\documentclass[11pt, tightenlines, onecolumn, amsmath, amssymb, notitlepage, a4paper, nofootinbib, superscriptaddress]{revtex4-1}

\usepackage[strict]{revquantum}

\usepackage{graphicx}
\usepackage{subfig}
\usepackage{braket}
\usepackage[margin=1in]{geometry}
\usepackage{bbm}
\usepackage{enumitem}
\usepackage[normalem]{ulem}

\DeclareMathOperator{\tr}{tr}

\newcommand{\nocontentsline}[3]{}
\newcommand{\tocless}[2]{\bgroup\let\addcontentsline=\nocontentsline#1{#2}\egroup}

\usepackage{thmtools}
\usepackage{thm-restate}

\begin{document}

\title{\Large Modeling and Control of a Reconfigurable Photonic Circuit using Deep Learning}
\author{Akram Youssry}
\affiliation{University of Technology Sydney,
		Centre for Quantum Software and Information,
		Ultimo NSW 2007, Australia}
\affiliation{Department of Electronics and Communication Engineering, Faculty of Engineering, Ain Shams University, Cairo, Egypt} 

\author{Robert J. Chapman}
\affiliation{Quantum Photonics Laboratory and Centre for Quantum Computation and Communication Technology, School of Engineering, RMIT University, Melbourne, Victoria 3000, Australia}
\affiliation{Institut f\"ur Experimentalphysik, Universit\"at Innsbruck, Technikerstra\ss e 25, 6020 Innsbruck, Austria}

\author{Alberto Peruzzo}
\affiliation{Quantum Photonics Laboratory and Centre for Quantum Computation and Communication Technology, School of Engineering, RMIT University, Melbourne, Victoria 3000, Australia}

\author{Christopher Ferrie}
\affiliation{University of Technology Sydney,
		Centre for Quantum Software and Information,
		Ultimo NSW 2007, Australia}

\author{Marco Tomamichel}
\affiliation{University of Technology Sydney,
		Centre for Quantum Software and Information,
		Ultimo NSW 2007, Australia}

\date{\today}

\begin{abstract}
The complexity of experimental quantum information processing devices is increasing rapidly, requiring new approaches to control them. In this paper, we address the problems of practically modeling and controlling an integrated optical waveguide array chip---a technology expected to have many applications in telecommunications and optical quantum information processing. This photonic circuit can be electrically reconfigured, but only the output optical signal can be monitored. As a result, the conventional control methods cannot be naively applied. Characterizing such a chip is challenging for three reasons. First, there are uncertainties associated with the Hamiltonian model describing the chip. Second, we expect distortions of the control voltages caused by the chip's electrical response, which cannot be directly observed. And third, there are imperfections in the measurements caused by losses from coupling the chip externally to optical fibers. We have developed a deep neural network approach to solve these problems. The architecture is designed specifically to overcome the aforementioned challenges using a Gated Recurrent Unit (GRU)-based network as the central component. The Hamiltonian is estimated as a blackbox, while the rules of quantum mechanics such as state evolution is embedded in the structure as a whitebox. 
The resulting overall graybox model of the chip shows good performance both quantitatively in terms of the mean square error and qualitatively in terms of the shape of the predicted waveforms. We use this neural network to solve a classical and a quantum control problem. In the classical application we find a control sequence to approximately realize a time-dependent output power distribution. For the quantum application we obtain the control voltages to realize a target set of quantum gates. 
The method we propose is generic and can be applied to other systems that can only be probed indirectly.
\end{abstract}

\maketitle
\tableofcontents

\section{Introduction}

The complexity of experimental quantum information processing devices is increasing rapidly, requiring new approaches to control them. Noisy Intermediate-Scale Quantum (NISQ) devices are emerging nowadays, with lots of experimental challenges \cite{preskill2018quantum, friis2018observation, arute2019quantum}. In this present work, we deal with the problem of modeling a device that can process some input signals to generate output signals, and the operation of the device can be manipulated using control signals. There are three possible methods to model such a device presented as follows.

The first approach is through direct physical modeling. We look for a mathematical description of the output signals expressed in terms of the input and control signals. The equations will involve some unknown parameters which should be chosen to match the performance of an actual realization of the device. And thus we perform measurements on the device and use methods of parameter estimation in order to find the unknown parameters of the model. We call this approach a whitebox approach. This would be the first approach one would try to use. The problem however is that if there are uncertainties in the relations between some variables, or some assumptions are made to derive some formulas (which might not be true for an actual device), then the resulting model might not be accurate enough to fit and predict actual measurements. Imperfections in the measurement process will also decrease the accuracy of the obtained model. Additionally, relations between some variables may be completely unknown and thus the problem becomes not just estimating parameters but also estimating functional forms (maps between variables). Moreover, there are situations where estimating the unknown parameters requires measurements that are not experimentally possible. For instance, if we want to estimate the parameters of a transfer function of an electrical circuit, then we will need to measure voltages at some nodes of the circuit. However, if we cannot physically access those nodes then the problem becomes more difficult. Finally, the complexity of the problem increases if the physical models involve non-linear relations. Thus, the whitebox approach might face lots of challenges in practical situations.

The second way to solve the problem is through the blackbox approach. We do not obtain a set of physical equations describing the device, but rather we construct a generic function that approximates the relationship between the output and the input and control signals. This is usually a highly non-linear function with a large number of parameters that can be estimated using the measurements. If the function is complex enough, then it can model and predict any unknown relations between variables. For that type of modeling, machine learning structures, such as artificial neural networks, are very suitable. This approach has an advantage of being capable of predicting the output signals given the input and control signals. However, there are few drawbacks. First, the resulting model provides the least amount of information about the physics of how the device works. And so it would be difficult to use the model for re-engineering the device if required. Second, the resulting accuracy may not be as high as expected. This is because the structure does not have any prior information about the map between inputs and outputs, and so it might need to ``discover'' some complicated laws of physics (such as the evolution of quantum systems) beside other unknown relations. This makes the training process harder. Consequently, a larger dataset and higher number of iterations would be needed to reach a good level of accuracy, which might turn out to be impractical.

The last approach is a combination of the other two approaches, we would refer to as a graybox model. In this case, we use direct physical modeling for parts of the description that we have complete certainty about (whitebox part), while we use a blackbox for the other parts that we are uncertain about. The model should be built such that the measurements required for the learning process are physically available; there is enough physical modeling through the whiteboxes to allow extracting useful information about the behaviour of the device; and any measurement imperfections should be accounted for. Machine learning structures are also suitable for this type of modeling. Standard machine learning layers would be used for the blackboxes. However, we also need to define non-standard layers to account for the whiteboxes, and these are application specific. The overall structure should be consistent to allow standard learning algorithms to work. In this paper, we explore the use of hybrid deep learning architecture to solve problems related to experimental modeling and control of quantum systems. Although the focus is on a photonic device that will be introduced shortly, the graybox approach can be considered very general, applying to many situations where there is a system that cannot be probed arbitrarily as discussed.

An example of this approach is when we have a quantum device described by a quantum system. The input signal is modeled by the initial quantum state. The output signal is modeled by a measurement performed on the system after evolving according to a given Hamiltonian. The control signals would then be some external forces applied to the system, and they are modeled by some terms in the Hamiltonian. Using the laws of quantum mechanics we could write down the relation between the input, output, and control. This would correspond to the whitebox part of the model. Some of the terms inside the Hamiltonian might be unknown, so we would use a blackbox to evaluate these terms. The resulting overall model is then a graybox model. This is a useful approach because it still gives an insight on the physics of the device and one can evaluate physically significant quantities. Additionally, the terms that reduce the accuracy of the models, due to inaccurate physical modeling, are now replaced by blackboxes, resulting in a more accurate overall model.

\begin{figure*}[t]
\centering
\includegraphics[width=\textwidth]{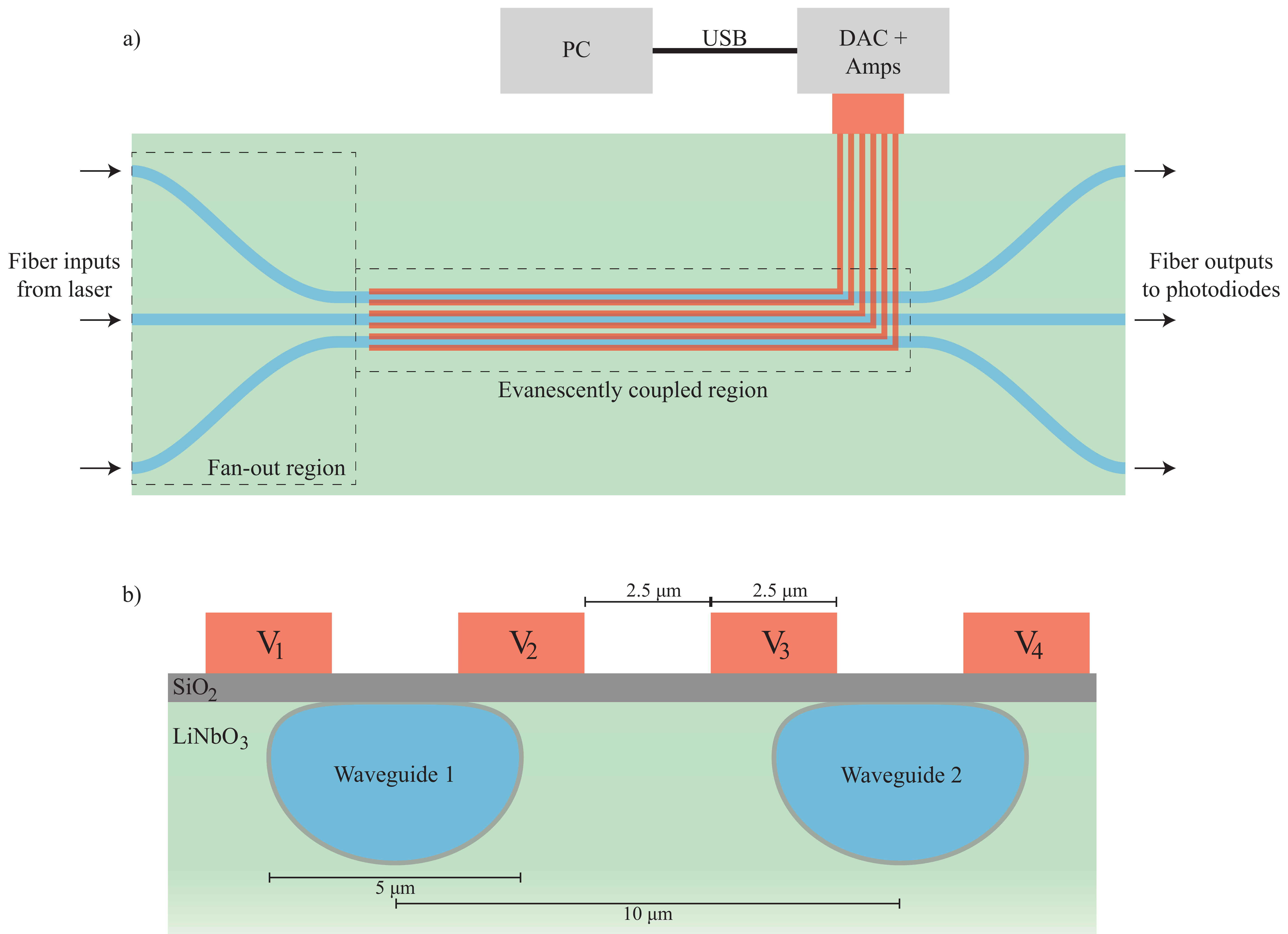}
\caption {a) Top view and b) cross section schematics of a three waveguide reconfigurable array. The waveguides initially fan-in from 127$\mu$m spacing to enable coupling to optical fibers. The waveguides in the array are separated by 10$\mu$m which enables nearest-neighbor evanescent coupling. The electric field between the electrodes causes a local change in refractive index to the waveguide or the cladding.}
\label{fig:chip}
\end{figure*}

In this paper, we focus on a particular system, currently being developed by some of the authors, which is an array of nearest neighbor coupled waveguides with a reconfigurable Hamiltonian. Characterizing such a chip is a significant challenge as will be discussed later. The device we consider is an array of nearest neighbor coupled waveguides that implements a continuous time quantum walk on photons propagating along the array \cite{PhysRevA.48.1687, peruzzo2010quantum}. In all previous work, static quantum walks were studied with fixed coupling parameters. Here, we demonstrate a reconfigurable waveguide array by exploiting the electro-optic control of Lithium Niobate. The waveguides are fabricated by reverse proton exchange and we apply local electric fields to change the properties of the coupled array. Figure \ref{fig:chip} shows the schematic of the chip. We inject laser light into one input waveguide of the array and measure the output optical power distribution across all the waveguides. The electrodes can be controlled to alter the output distribution.

Numerical simulations of such a device shows a host of potential applications. The chip can operate as a classical device with possible applications in telecommunications such implementing a Mach-Zehnder interferometer or an electro-optic modulator. Being able to characterize and control such a device is important and has a strong economic impact, but at the same time is very challenging as will be discussed later. Additionally, the chip can work as a quantum device. This includes operating as a quantum router, where single photons can be directed to propagate and be detected at one of the output ports by dynamically changing the control voltages. It can also be used to generate complicated quantum states (such as W state), and realize different quantum gates on single or multiple photons. 

We focus on two different scenarios of using this device. The first is when experimentally we are only measuring powers at the output of the chip. This is also equivalent to a single-photon experiment where we only detect photons at the outputs. In this situation, we can use classical modeling only. However, we use a quantum model for two reasons. First, to show the applicability of the graybox approach when the whitebox parts are quantum. The second reason is that although we only measure powers at the output, we allow for arbitrary states in the input, including entangled states which cannot be described by a pure classical model. Also, if there are multiple input photons at different waveguides, then a quantum mechanical description of the chip is required to describe the correlations of the photons at the output \cite{peruzzo2010quantum}. The second application is when we can also measure phases through Mach-Zehnder type of interferometry. In this case, we show the possibility of implementing single-qubit quantum gates with high fidelity, where the qubits live in the subspace formed of the two far-end waveguides. The proposed framework allows finding the set of control voltages required to obtain a target sequence of quantum gates, given the different challenges faced during the characterization process.

Machine learning has been a very active area of research recently, with focus on both the algorithms as well as the wide range of applications touching every field of science and beyond. Deep learning has particularly gained attention as it becomes more and more feasible. This is due to today's enormous computational power, as well as the availability of big datasets for training. The survey \cite{deng2014tutorial} covers the common architectures used in deep learning and the range of possible applications. 

The physics community is also currently exploring the use machine learning to solve some practical problems faced in designing, controlling, and automating experiments. Some examples of recent work include the automated design of quantum optical setups \cite{krenn2016automated} using reinforcement learning \cite{melnikov2018active}, and using deep learning and genetic algorithms \cite{o2018hybrid}. Machine learning was also used in \cite{zhou2019self} to configure an optical signal processor, which itself can work as an artificial neural network with linear activation functions. Deep learning was also used in Ref. \cite{ming2018quantum} to discover and characterize topological phases of matter and phase transitions. Techniques of both deep learning and reinforcement learning have been applied in quantum control \cite{niu2019universal, PhysRevX.8.031086,ostaszewski2019approximation}. These works differ from ours by treating the entire learned model, including quantum dynamics, as a blackbox, with no detailed modeling of an experimental realization.

Methods of machine learning have also been used in other areas of quantum information. For example, the work presented in \cite{aaronson2018online,youssry2019efficient} is about developing online quantum state estimation algorithms inspired by the matrix exponentiated gradient method, a technique used in classical machine learning. Other applications include the use of neural networks in quantum cryptography \cite{Niemiec2019} and in quantum error correction \cite{baireuther2018machine}. Another related problem to what we present here is Hamiltonian learning \cite{granade2012robust, wiebe2014hamiltonian, wiebe2014quantum, wang2017experimental}. This is a Bayesian framework that allows updating the priors on Hamiltonian estimates given observed measurements. This approach although very useful on its own, is not suitable for the problem under consideration. The reason is that we are interested in estimating the map between control voltages and the Hamiltonian through indirect measurements. The Bayesian approach is suitable if the Hamiltonian is fixed (the control voltages are fixed). Another Bayesian approach is presented in \cite{lennon2019efficiently} where the focus is on the real-time prediction of the set of optimal measurements to perform on a quantum dot, using partial information available so far. This allows efficient characterization of the device.

The structure of the remainder of the paper is as follows.
The paper starts with an overview on the quantum-mechanical description of the chip in Section \ref{sec:chip}, followed by the experimental constraints and challenges in Section \ref{sec:challenges}. Next, in Section \ref{sec:methods} we present the proposed deep learning architecture in detail. After that, we present the numerical results of the simulations and discuss their significance in Section \ref{sec:results}. Finally, we end with the conclusion and discuss the possible future extensions of this work in Section \ref{sec:conclusions}. Appendix \ref{sec:suppfig} contains figures related to Section \ref{sec:results} placed there for maintaining the readability and continuity of the text.
\section{Problem Setup}

This section starts with describing quantum mechanically the photonic circuit we are trying to model and control, followed by the challenges we face in characterizing it experimentally.

\subsection{Chip model}\label{sec:chip}

The chip with $n$-waveguides can be described quantum mechanically in $\mathbb{C}^n$ Hilbert space, with the computational basis encoding the presence of photons in each waveguide. For example for $n=3$ the state $\ket{0}=[1,0,0]^T$ encodes a photon present at the first waveguide, the state $\ket{1}=[0,1,0]^T$ encodes a photon in the second waveguide and so on. The evolution of the system represents the behavior of the chip when light propagates along the waveguides. So, the initial state of the system represents the mode distribution at the inputs of the waveguides, while the final state represents the distribution at the output of the waveguides. For example, if the system evolves from the the state $\ket{0}$ to the state $\ket{1}$, then this means that we started with injecting a photon at the first waveguide (at one end of the chip), and the photon got perfectly transfered to the second waveguide after propagating along the chip until the output. This evolution can be described by the unitary 
\begin{align}
U = e^{-i H l},
\label{equ:U}
\end{align} 
where $l$ is the length of the chip, and $H$ is the Hamiltonian of the chip. In general, we can write the Hamiltonian in the form
\begin{align}
H = H_0 + H_I(\bf{v}),
\end{align}
where $H_0$ is the zero-voltage Hamiltonian, and $H_I$ is the interaction Hamiltonian which is a function of the voltages $\bf{v}$ applied on the electrodes. Note that the control voltages are time-dependent, however, the time scale of the change is much slower than the time scale of the photon travel across the chip. That is, each photon can see only one time-independent Hamiltonian from the moment it enters the chip until the moment it reaches the output. But the next photon to arrive can experience a different Hamiltonian. This assumption is plausible since it is impossible to change the voltage faster than the flight time of the photon in the chip. This is what allows us to write the evolution as the matrix exponential of the Hamiltonian as in Equation \ref{equ:U}, without the time-ordering operator.

\begin{figure*}[t]
\includegraphics[width=0.5\textwidth]{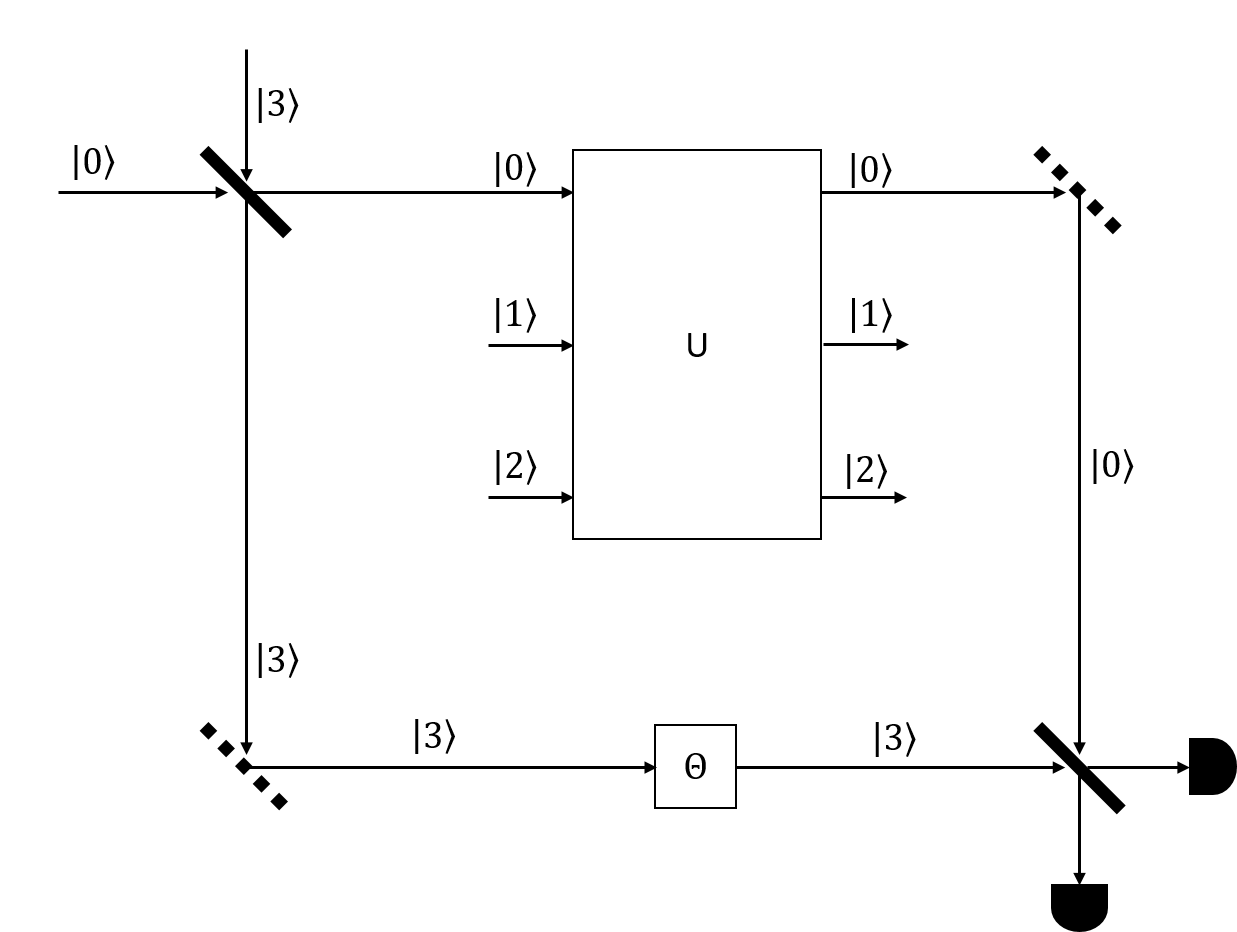}
\caption{A schematic for a typical Mach-Zehnder interferometer to measure the phase shift between the inputs and outputs of the photonic chip. The kets on the optical paths represent the encoding along this direction. The dotted lines are mirrors, while the solid thick lines are beamsplitters.} 
\label{fig:machzehnder}
\end{figure*}

In the basic experimental setup we can only measure output power distribution. For example, for an $n=3$ chip, if the input state is $\ket{0}$, and the output state after evolution is $U\ket{0}=\alpha\ket{0} + \beta \ket{1}+\gamma \ket{2}$, then the output distribution we measure is $(|\alpha|^2,|\beta|^2,|\gamma|^2)$. However, to characterize a fully quantum model, we need to measure phases at the output. One of the convenient ways experimentally to measure relative phase shifts between two optical paths is through Mach-Zehnder interferometery as shown in Figure \ref{fig:machzehnder}. Recall the basic idea is to construct a quantum circuit whose output probability amplitude depends on the phase shift required to be measured. With an initial state $\ket{0}$, a standard calculation shows that the final state after the beamsplitter at the bottom-right of the diagram is
\begin{align}
	\ket{\psi}=\frac{1}{\sqrt{2}}\left(\frac{\alpha + e^{i\theta}}{\sqrt{2}}\ket{0} + \beta\ket{1} + \gamma\ket{2} + \frac{\alpha - e^{i\theta}}{\sqrt{2}}\ket{3} \right)
\end{align}
Now, if we measure the power at the detector, we get $P(\theta)= \frac{1}{4}|\alpha+e^{i\theta}|^2$. Now if we do two of such measurements corresponding to values of $\theta=0$ and $\theta=\frac{\pi}{2}$, we can exactly calculate both the amplitude and phase of this output. Particularly, 
\begin{align}
	|\alpha|^2+1+2|\alpha|\cos{\angle\alpha}&=4P(0)\\
	|\alpha|^2+1+2|\alpha|\sin{\angle\alpha}&=4P\left(\frac{\pi}{2}\right),
\end{align}
where $\angle\alpha$ denotes the phase of $\alpha$. These two equations can be solved simultaneously to find the amplitude and phase of $\alpha$. Now, the procedure can be repeated by placing the mirror at the top-right of the diagram at all other outputs of the chip and obtain the amplitude and phase of this part of the state. Since we have an $n$-dimensional pure state, it is completely defined by $2n$ degrees of freedom corresponding to real and imaginary part of each coefficient. (In fact, only $2n-2$ are needed since we have the normalization constraint, and a non-significant global phase shift). The same procedure can be executed to characterize the output state when other inputs are activated. Finally, it is worth mentioning that this setup for measuring phase is not the only possible way, there might be a more efficient way to measure the phases at the output without requiring to move the optical components spatially. This is however out of the scope of this paper. 

\subsection{Experimental challenges}\label{sec:challenges}
There are many experimental challenges faced when characterizing a fabricated chip, as well as designing the control voltages to implement some desired behavior. Any model for the device should account for these constraints. These challenges are listed as follows.

\begin{enumerate}
\item The drifting in the measured output optical power.

This problem is caused by charges getting trapped at the interface between the Silicon Dioxide and the Lithium Niobate\cite{yamada1981dc,gee1985minimizing,nagata1995progress}. These charges have very low mobility and therefore take a long time to accumulate and a long time to diffuse when the voltage is removed. These trapped charges are the central reason we have difficultly controlling and characterizing this device. The long diffusion time results in the voltage never `resetting' to zero. It then becomes extremely difficult to infer what electric field is being applied to the waveguide. In any case, the chip has some equivalent electrical circuit model. But this is difficult to model and characterize experimentally, as we cannot measure physically the voltages the chip actually sense when we apply externally some control voltages. The only available measurements are the output waveguide power distribution, which depends non-linearly on the control voltages. This makes the problem a non-linear control and estimation problem and that is classically difficult to solve. These effects cannot be neglected as well because the distortions in the control voltages will be reflected on the measured power distribution. It will also have a memory effect in the sense that when we apply some control pulse, the output power will be affected by that pulse in addition to the previous pulses that were applied. This means that if at some point in time we set all the control voltages to ground, we will still observe variation of the power distribution in time. The classic way of overcoming this problem is during fabrication by etching the buffer layer between the electrodes \cite{yamada1981dc}. However, for the particular chip we are working with, the dimensions are very small and technologically it is difficult to do this process. Thus, this problem has to be addressed differently. Therefore, the model should account for these unknown distortions, and it should be trainable using only available power measurements.

\item The uncertainties regarding the structure of the Hamiltonian.

Usually, the Hamiltonian in these chips is assumed to have a tridiagonal form reflecting the fact that only adjacent waveguides are coupled \cite{bromberg2009quantum, peruzzo2010quantum}. But there is a possibility that there are non-negligible higher order couplings between the waveguides leading to more off-diagonal terms. Also, one could assume the linear dependence of the Hamiltonian on the control voltages. But, this assumption might not be true as there might be higher order terms. Thus, the model should not assume any particular form of the Hamiltonian except that it is Hermitian as required by quantum mechanics.

\item The power losses at the output.

Losses in the measured power occur due to the coupling of the chip to the external optical fibers connected to the photodetector. These will cause inaccuracies in the measurements affecting any parameter estimations. These losses also have to be characterized so that we can make corrections for the detected power signals. We will model the losses by
\begin{align}
\hat{P}_k = \frac{\epsilon_k P_k}{\sum_{i=1}^n{\epsilon_i P_i}}, 
\end{align}
where $\hat{P}_k$ is the $k^{th}$ normalized measured power at waveguide $k$, and $P_i$ is the actual power at the output of the chip for waveguide $i$. The normalization is for making the measurements constitute a probability distribution. The model should account for these losses.

\item The limitation on the control voltages.

Generally, in order to obtain some target output for the device, we need control voltages that can be arbitrarily large. However, if the potential difference across any pair of electrodes exceeds some maximum value, the device will break down. It might be the case that within this limitation one cannot obtain the target with infinite precision. This controllability issue is a different problem and is a subject of the future work of this paper. And so, any control algorithm should try to maximize the accuracy of the target output without exceeding the allowed range for the control voltages.
\end{enumerate}

As a result of all the previous challenges, estimating and controlling the Hamiltonian directly from measured data is very difficult using the whitebox approach. The complete blackbox might perform well but as discussed it will not give physical insights on the device. Thus, we will seek the graybox approach for modeling the chip under the aforementioned constraints. The blackbox part will represent the map between the Hamiltonian and the control voltages. This allows getting rid of any assumptions on the Hamiltonian as well as accounting for the pulse distortions. The whitebox part will represent the other certain relations derived from quantum mechanics. The next section will give more details about how to construct such a model using deep learning.


\section{Methods}\label{sec:methods}
In the previous section we described the challenges we face in experimentally characterizing the chip if we use conventional methods of model and parameter estimation. In order to address all these challenges, we propose to use a completely data-driven approach rather than a parametric approach. We are going to use graybox model where the Hamiltonian will be treated as a blackbox, while the quantum evolution and quantum measurement will be treated as whitebox. This is because all the uncertainties are in the Hamiltonian, while the all the laws of quantum mechanics are known. We will design a deep learning structure to implement this idea. The problem will be divided into two stages. The first stage, a set of known control voltages and corresponding power distribution will be used by a supervised deep learning algorithm to find a complete graybox model for the chip. The second stage will be creating another deep learning structure to find the control voltages that results in some desired behavior of the chip, using the estimated model from the first stage. 

This section starts with a detailed description of the architecture used to model the chip. Next, the training and testing procedures are presented. After that, the detailed description of the control voltages predictor for the chip is presented. Finally, the section ends with extending the proposed structure to account for a fully-quantum setting where phases can be measured at the output.

\subsection{Chip model architecture}
\begin{figure*}[t]
\centering
\includegraphics[width=0.75\textwidth]{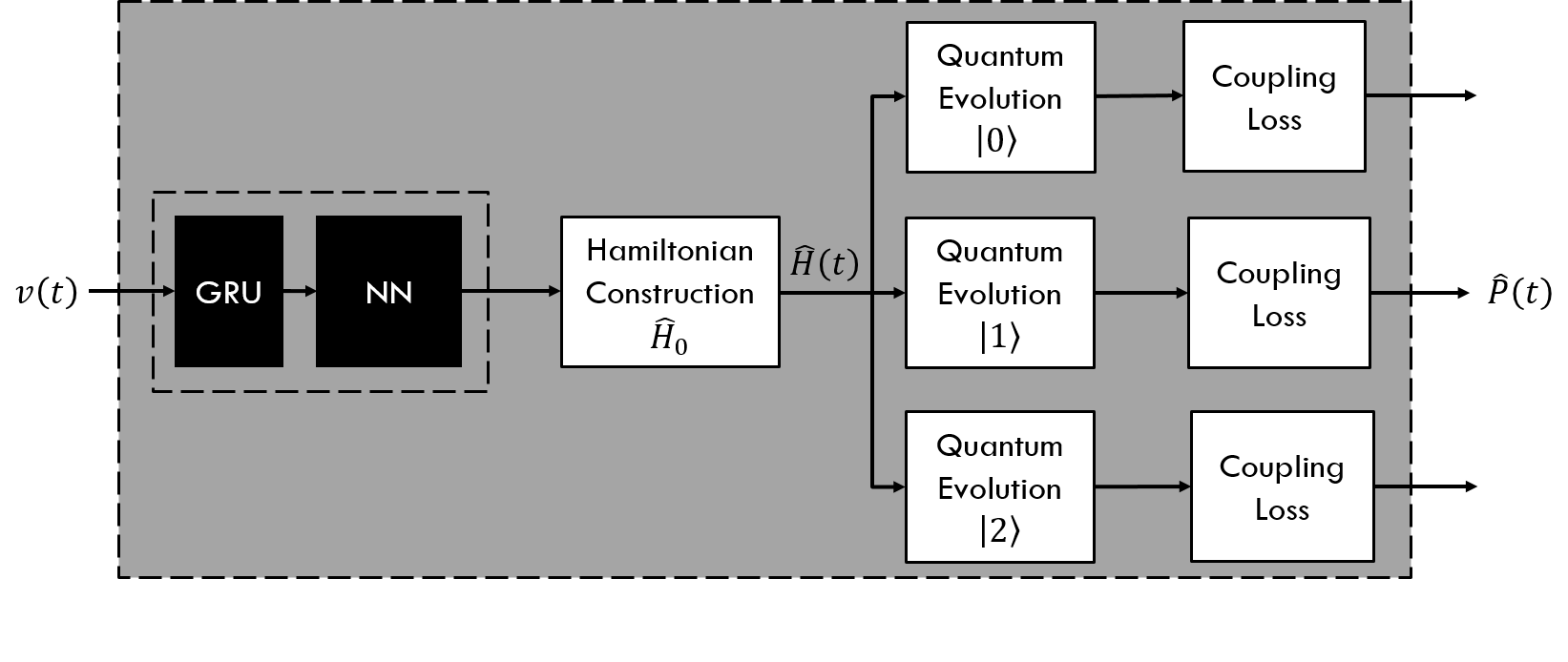}
\caption {The different layers for the proposed graybox architecture to model a 3-waveguide chip ($n=3$), with inputs representing the control voltages and outputs representing the measured optical power. The first two layers together represent the blackbox part of the model, while all other layers represent the whitebox part. The Gated Recurrent Unit ``GRU'' layer together with the Neural Network ``NN'' layer model the map between the interaction Hamiltonian and the control voltages, as well as the distortions of the control voltages. The ``Hamiltonian construction'' layer converts the output from the previous layer into a Hermitian matrix, and adds the zero-voltage Hamiltonian. The ``Quantum Evolution'' layer calculates the the probability amplitudes of the evolved quantum state given the Hamiltonian from the previous layer. The ``Coupling Loss'' layer models the power loss during measurement. The branching ensures that we get the same Hamiltonian for different initial states to be consistent with linearity of quantum mechanics. The graybox model is trained over two stages. The first stage is training the whitebox parts using zero-voltage measurements. The second stage is training the blackbox parts using random pulse measurements.} 
\label{fig:model}
\end{figure*}

The deep learning architecture is shown in Figure \ref{fig:model}. The first layer in the model is a Gated Recurrent Unit (GRU) \cite{cho2014learning}. This is a variant of the Long-Short Term Memory (LSTM) structure often used in sequence prediction and classification \cite{hochreiter1997long}. GRU is more efficient than LSTM as it has fewer parameters to be learned during the training stage. However, in terms of accuracy, it is not very clear which is better generally, and this remains an open topic under investigation within the machine learning community \cite{chung2014empirical}. The number of inputs is equal to the number of electrodes which is $2n$. For our implementation, the number of hidden units of the GRU is chosen to be 60. In general, more hidden units allow modeling more complex waveforms, but on the expense of more parameters to learn and thus more computational resources required. The objective of this layer is to learn the interaction Hamiltonian, i.e. learn how the Hamiltonian depends on the external voltages. This should also include the parasitic effects in the chip causing distortions of the applied voltage waveforms. The number of free parameters of any real-valued symmetric Hamiltonian of size $n\times n$ is $\frac{n}{2}\left(n+1\right)$. However, the output of the GRU is the output of the each hidden node. So, to extract the required number of outputs, we add a neural network (NN) formed of a single layer that is fully-connected to all of the outputs of the GRU. The number of neurons is exactly equal to $\frac{n}{2}\left(n+1\right)$, as each neuron generates one output. Linear activation is used for all neurons, to allow the output to take any value and not be restricted in some range if we use other activations such as sigmoid. Notice, that the GRU is a sequential layer, so the output has an extra dimension of time. However, the NN layer is static acting equivalently on each time slice of the output of the GRU. This means that weights applied to the GRU output at every time instant are the same. These two layers together act as a device to learn the free parameters of the Hamiltonian as a function of the input voltages.

The third layer in the structure is a custom-defined layer that has two functionalities. The first one is to reconstruct a symmetric matrix from the output of the previous layer. This is done by reshaping the outputs as an upper triangular matrix, and then sum it with its transpose. The second functionality is to add to the drifting Hamiltonian, that is the zero-voltage Hamiltonian that models the inherent coupling between the waveguides. The parameters of this drifting Hamiltonian are learned during the training process as will be illustrated later. The final output of this layer is therefore the full Hamiltonian of the system.

The next layer of the model is the quantum evolution layer. This is a custom defined layer, that takes some Hamiltonian as input, an initial quantum state as a defining parameter, and generates the probability amplitudes of the an evolved state as output. These probability amplitudes correspond to the waveguide power distribution. So, the layer first calculates the evolution matrix $U = e^{-i H l}$. Next, it calculates the evolved state $\ket{\psi_F} = U\ket{\psi_0}$. Finally, it calculates the probability amplitudes of the evolved state $|\braket{m|\psi_F}|^2, m=0,1,\ldots,n-1$.

Now, a problem arises if we train the model with the structure so far. Since, only one initial state is used in the quantum layer, then the learned Hamiltonian will be valid only for evolutions of this state. But, if we use the same Hamiltonian to evolve other initial states, we might not obtain a correct evolution. So, the algorithm will need to learn a different Hamiltonian for each initial state. This is a major problem, since quantum mechanics is a linear theory, so the Hamiltonian should not depend on the quantum state being evolved. Thus, we have to constrain the Hamiltonian in some sense so that it works for all states. The way we propose to solve this problem is to have different copies of the quantum layer each parameterized by a different initial state. Then, we connect the input of all these layers to the same output of the previous Hamiltonian layer. In this case, during the training, the model will be enforced to generate a Hamiltonian that correctly evolves each of the initial states. Since a unitary can be completely characterized by knowing the outputs corresponding to each of the computational bases as input states, we only need $n$ of `parallel' quantum structures each generating $n$ outputs. So, the total number of outputs for this whole layer is $n^2$. 

The final layer in the model is also a custom-defined layer that models losses during power measurements. This physically occurs due to coupling between between the chip and optical fibres connected to the photodetectors.  The layer simply implements the calculation $\hat{P}_k = \frac{\epsilon_k P_k}{\sum_{i=1}^n{\epsilon_i P_i}}$, where $\hat{P}_k$ is the $k^{th}$ measured power at waveguide $k$, and $P_i$ is the actual power at the output of the chip for waveguide $i$. The denominator in the expression is to ensure that the measured powers are normalized, (i.e. form a distribution). The coupling coefficients are learned during the training stage as will be discussed later. For each quantum block in the quantum evolution layer, we cascade one of these coupling layers. However, all of these copies of the coupling layers are identical (i.e. have the same parameters). This reflects the fact that the losses are independent of which waveguide was used as input, and just related to the hardware of the experiment. 

\subsection{Training and Testing}
There are two stages to do the training of the model, where all the unknown parameters of the model are leaned by providing examples. The first stage is to learn all the zero-voltage parameters, i.e. the drifting Hamiltonian and the coupling losses coefficients. All these parameters are static and do not depend on the input voltages. For this training step, we detach the GRU and NN layers from the model. The input of the model is directly connected to the Hamiltonian construction layer, and is fixed to be all zeros. The output is the lossy power distribution. This is obtained experimentally by fixing the physical voltage on the chip to zero, using one of the waveguides as input and measure the power across each waveguide. The procedure is repeated for all input waveguides. Since, the distribution in this case is static, we get a total of $n^2$ readings. With this pair of training data (zero voltage as input, and $n^2$ readings as output), the model is trained by backpropagation using RMSprop \cite{tieleman2012lecture}, and all the unknown parameters are learned. We use the mean square error (MSE) as the loss function and also as the performance metric. This is because the problem is predicting a waveform, and MSE is one of the most commonly-used metrics for quantifying similarity between two waveforms. The lack of phase information at the output prevents us from constructing a full quantum state and thus evaluating quantum measures such as fidelity is not possible.

The second stage of training is to obtain the dynamic behavior of the chip, (i.e. how the waveguide power distribution changes in time being a function of the input time-varying voltage. In this stage, the full model is used, and the input is connected to the GRU layer. All parameters learned from the first stage are fixed and do not change during this stage. Backprogation is used to train the remaining unknown parameters using the pair of some voltage waveforms as input, and the corresponding measured power distribution waveforms as output, with MSE acting as loss function. After this stage, all the learned parameters are fixed and the model can be used in the testing phase.

In the testing phase, the model is given a new input that was not in the training set, and the predicted output is compared with the actual output. A good model is a one that generalizes well over new inputs. The end goal of using this architecture is a graybox model of the chip, capable of predicting the output distribution for any control voltage. However, practically this is a hard requirement due to the behavior of machine learning algorithms. Usually, these structures have the ability to generalize for inputs that share some similarity with the training examples. In our case, the voltage waveform shape should be the same for the training and testing datasets (i.e. fix the pulse shapes to be either square, Gaussian, raised cosine...etc.). After fixing the shape, the waveform parameters (such as amplitude, phase shift,...etc.) for each example can be arbitrary. If we want the model to predict the output for other waveform shapes, then the training set has to include the other shapes as well.  In this paper, we restrict all the voltage waveforms to have the form of arbitrary synchronized square pulses. This means that for each example, the pulses across all electrodes start at the same time instant, have the same width, but can have different amplitudes. These parameters will differ though across different examples in the datasets.

The architecture of this model has a major advantage which is the possibility of monitoring the output of each layer during testing, each corresponds to a physically significant quantity. So, the output of the NN layer is a prediction of the interaction Hamiltonian as a function of the input voltages and time. The output of the ``Hamiltonain Construction'' layer is a prediction of the total Hamiltonian matrix. The ``Quantum Evolution'' layers predict the ideal power distribution for each initial state, while the output of the last layer is prediction of the measured power distribution. This shows that relevance of this deep learning structure. For instance, had we used one LSTM-based blackbox instead of the proposed graybox, we would have been able to predict the measured power distribution only, but not the other quantities.

\subsection{Controller Architecture}
\begin{figure*}[t]
\centering
\includegraphics[width=0.75\textwidth]{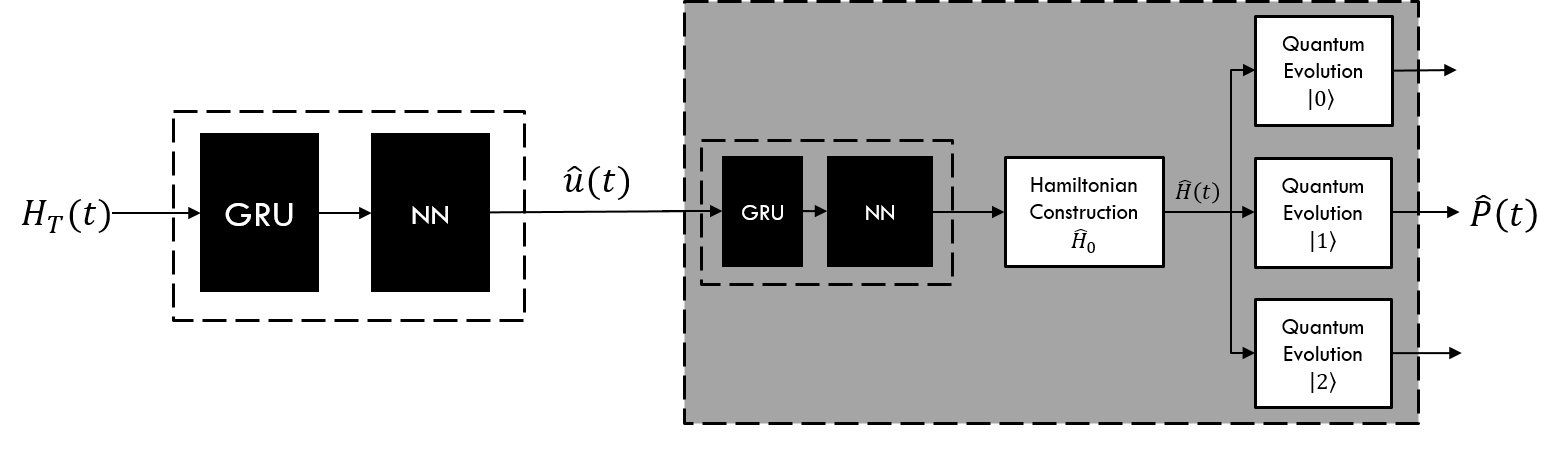}
\caption {The proposed architecture of the controller for the $n=3$ chip, with inputs representing the target sequence of Hamiltonians and outputs representing the target controlled optical power distribution. The trained graybox model of the chip is fixed preventing any further changes. An extra blackbox is added at the beginning. The Gated Recurrent Unit ``GRU'' and the Neural Network ``NN'' layers model the map between the sequence of target Hamiltonians and control voltages. The activation function in the NN layer is chosen to ensure that the resulting control voltages are limited to the allowed range the chip supports. After training, the control voltages can be estimated by probing the output of the ``NN'' layer.}
\label{fig:controller}
\end{figure*}

The second major task required after characterizing the chip is finding the control voltages needed to obtain a desired power distribution, resulting from the evolution of a target Hamiltonian. The architecture of prosed controller is shown in Figure \ref{fig:controller}. The first layer is a GRU layer followed by a fully-connected neuron layer similar to that used in the model architecture. However, the input is some desired target Hamiltonian, and output shall represent the control voltages which is a $2n$ vector. Since we need at least one of the electrodes to be connected to ground, we actual enforce the very first electrode to zero. Also, we enforce the last electrode arbitrarily to zero. This leaves out $2n-2$ control voltages to predict. For efficiency purposes, we actually input only the upper triangular part of the Hamiltonian flattened into an $\frac{n}{2}\left(n+1\right)$ vector. 

One major issue to consider is that the voltage across any two adjacent electrodes should not exceed in absolute value $V_{\max}$. So, all the neurons at the output have a scaled hyperbolic tangent sigmoid activation in the form $f(x) = \frac{1}{2}V_{\max} \tanh(x)$. This ensures the output at each electrode is in $[-\frac{1}{2}V_{\max},\frac{1}{2}V_{\max}]$, and thus the potential difference across any two adjacent electrodes is limited to $[-V_{\max},V_{\max}]$. 

Next, we cascade a copy of the previously trained model without the coupling loss layers.  The reason behind dropping that layer is that the power loss is due to the measurement process, and not the operation of the chip. For instance, if two chips were connected in cascade with perfect coupling, then we would be interested to predict the control voltages for the first chip to produce some desired state at its output, and there will be no effects of the losses for the first chip. All the trained parameters of the model are fixed and do not change during the training of the controller. Connecting the pre-trained model enforces the whole controller structure to generate the ideal target power distribution. Thus, all the distortions that appear in the power distribution are dealt with automatically by the controller. The algorithm is enforced to produce voltage waveforms that undo the distortion effects in order to minimize the MSE. This means the algorithm is effectively learning an inverse model of the equivalent circuit of the chip, and simultaneously ensuring the final quantum state is correct. In some sense, this structure does both classical control (undoing the distortions) and quantum control (obtaining the target quantum state). By probing the output of the NN layer, the desired control voltage can be estimated.

It is worth mentioning that there is no requirement on the controller to generalize to every possible target Hamiltonian/target-distribution pair. Whenever we are interested to realize some sequence of operation on the chip, we redo the training of the controller, and probe the output of the NN layer. So, in some sense we are using backpropagation as a direct optimization procedure rather than a learning procedure. Additionally, the controller input is a sequence representing the Hamiltonian at each time step. This means we can obtain control voltages that allows changing the behavior of the chip dynamically whilst operating.

The last point to note is that not every possible Hamiltonian can be realized with the chip model. Some Hamiltonians may require voltages that exceed the maximum allowed range. An open question is what kind of quantum gates can be actually implemented using this chip given the constraints. This is however outside the scope of this paper.

\subsection{Fully-quantum model}
The architectures described so far are not fully quantum in the sense that the Hamiltonian is assumed to be real, and that we can only measure powers at the output (corresponding to probability amplitudes). However, it is possible to extend the proposed method to the fully quantum case, if we perform the Mach-Zehnder type of measurements as discussed previously. The overall architecture is quite similar, with the following modifications:
\begin{itemize}
\item The neural layer after the GRU is set to produce $n^2$ outputs instead of the $n(n+1)/2$, to account for the imaginary part of the Hamiltonian matrix elements.

\item The Hamiltonian layer reshapes the output of the neural layer to an $n\times n$ matrix, where the lower triangular part represents the imaginary part of the Hamiltonian while the upper triangular part represents the real part. So, by multiplying the lower triangular part by $i$ and adding the whole matrix to its Hermitian conjugate, we end up with an $n\times n$ Hermitian matrix. Also, the zero-voltage Hamiltonian $H_0$ is manipulated similarly to account for the possibility of complex-valued entries.

\item The quantum layer outputs the Mach-Zehnder interferometer power measurements instead of the probability amplitudes. So if the final state is $\sum_k{\alpha_k\ket{k}}$, then the layer's outputs are $P_k(0) = \frac{1}{4}|\alpha_k+1|^2$, and $P_k\left(\frac{\pi}{2}\right)= \frac{1}{4}|\alpha_k+i|^2$, for all $k=1,...n$. So, the total number of outputs for this layer is $2n$, and for the whole model is $2n^2$. We do not need to explicitly calculate the amplitude and phases from the interferometer measurements for the training. We will just use the interferometer measurements directly. The training follows the same procedure as discussed previously.

\item For simplicity, we removed the coupling layer as the focus in this application is on exploring the possibility of learning a fully quantum system. However, in general we can include it.

\item We still use MSE as a loss function and performance metric because the output is still a waveform (although representing interference measurements now). However, since there is complete information to reconstruct the state and the evolution matrix, we can use other metrics for performance evaluation such as fidelity.

\item The controller architecture is the same, the only difference is the input of the first layer is the real and imaginary parts of the target unitaries, rather than the Hamiltonians. This seemed to perform better than having the Hamiltonians as input. This might be due to the fact that there exist infinitely many Hamiltonians (all related with a factor of integer multiple of $2\pi$ in the eigenvalues) giving rise to the same unitary. And thus, the GRU might have trouble finding some of these equivalent Hamiltonians. However, if the input is directly the unitary then there is no redundancy. For the classical application, this did not seem to cause any problems because there was more freedom as the optimization is over the power distribution only. In the quantum application, it is more restricitve since the optimization is over the phase information as well.
\end{itemize}
\section{Simulation results}\label{sec:results}

This section discusses the implementation details of our method and the results of the numerical simulations. A discussion on the significance of the results is given afterwards. 

\subsection{Implementation}
For implementing the proposed architecture we used the ``Tensorflow'' Python package \cite{tensorflow}, and its high-level API package ``Keras'' \cite{keras}. The Python implementation of our algorithm is publicly available\footnote{https://github.com/akramyoussry/GRUBI}. 

In order to do training and testing, we created a dataset consisting of control voltages in the form of random pulses, and the corresponding waveguide output power distribution for different input waveguides. We generated a total of 4000 examples, 3500 of which were used for training and 500 for testing. The amplitudes of the pulses are from -5 to +5 volts and the time domain is limited to the interval $0 \le t \le 200 (ms)$ with sampling time of $0.2 (ms)$. In each example, the voltage on the first and last electrodes are fixed at zero, while the pulses are applied on the remaining electrodes. The restriction on these pulses is that they have to be synchronized across the different electrodes, starting and ending at the same time. However, the durations and amplitudes are chosen randomly from one example to another. The experimental setting would be generating these pulses, applying them physically to the chip, measuring the output power distribution, and finally training the model. However, in this paper, we restrict the study to computer simulations. So, we created a simulator for the chip that generates the waveguide power distribution given a set of control voltages, using the Hamiltonian model described by the tridiagonal real-valued matrix
\begin{align}
H = \begin{pmatrix}
\beta_1 & C_{1,2} & 0 & 0 & \cdots & 0\\ 
C_{1,2} & \beta_2 & C_{2,3} & 0 & \cdots &0\\ 
0 & C_{2,3} & \beta_3 & C_{3,4} & \cdots & 0\\
0 & 0 & C_{3,4} & \beta_4 & \ddots & \vdots\\
\vdots & \vdots & \vdots & \ddots & \ddots &  C_{n-1,n} \\
0 & 0 & 0 & 0 & C_{n-1,n} & \beta_n
\end{pmatrix},
\end{align}
where $\beta_i$ is the propagation constant along the $i^{\text{th}}$ waveguide, and $C_{i,j}$ is the coupling coefficient between waveguides $i$ and $j$. The propagation constant is given by
\begin{align}
\beta_i = \frac{2\pi}{\lambda}\left(n_0 + \Delta n \Delta V_i\right),
\label{equ:beta}
\end{align}
where $\lambda$ is the wavelength, $n_0$ is the intrinsic refractive index of the waveguide, $\Delta n$ is a dynamical proportionality constant that determines how much the the propagation constant changes by changing the voltage across the waveguide $\Delta V_i$. The coupling coefficient is given by
\begin{align}
C_{i,j} = C_0 + \Delta C_1 \Delta V_{i,j} + \Delta C_2 \left(\Delta V_i + \Delta V_j\right),
\label{equ:C}
\end{align}
where $C_0$ is the intrinsic coupling between two adjacent waveguides, $\Delta V_{i,j}$ is the potential difference across the substrate between the two waveguides $i$ and $j$, $\Delta V_i$ and $\Delta V_j$ are the voltages across waveguides $i$ and $j$, and $\Delta C_1$ and $\Delta C_2$ are dynamical proportionality constants that determine the amount of change of the coupling between two waveguides by changing the voltages across them. These relations assume that Hamiltonian depends on the voltages linearly, and that the coupling is always between neighboring waveguides. The simulator takes into account the non-ideal effects due to the equivalent circuit behavior of the chip, by simulating distortions on voltage pulses. It also simulates  coupling losses. For the results presented in this paper, the simulation parameters were as follows. $n=3$, $\lambda=808\times10^{-9}$, $l=3.6\times10^{-2}$, $n_0=2.1753$, $\Delta n =5\times10^{-6}$, $C_0=100$, $\Delta C_1=1.5$, $\Delta C_2 = -1.3$, and $\epsilon_k=\{0.9,0.8,0.5\}$.

\subsection{Results}
\begin{figure*}[h]
\centering
\subfloat[Classical]{\includegraphics[scale=0.75]{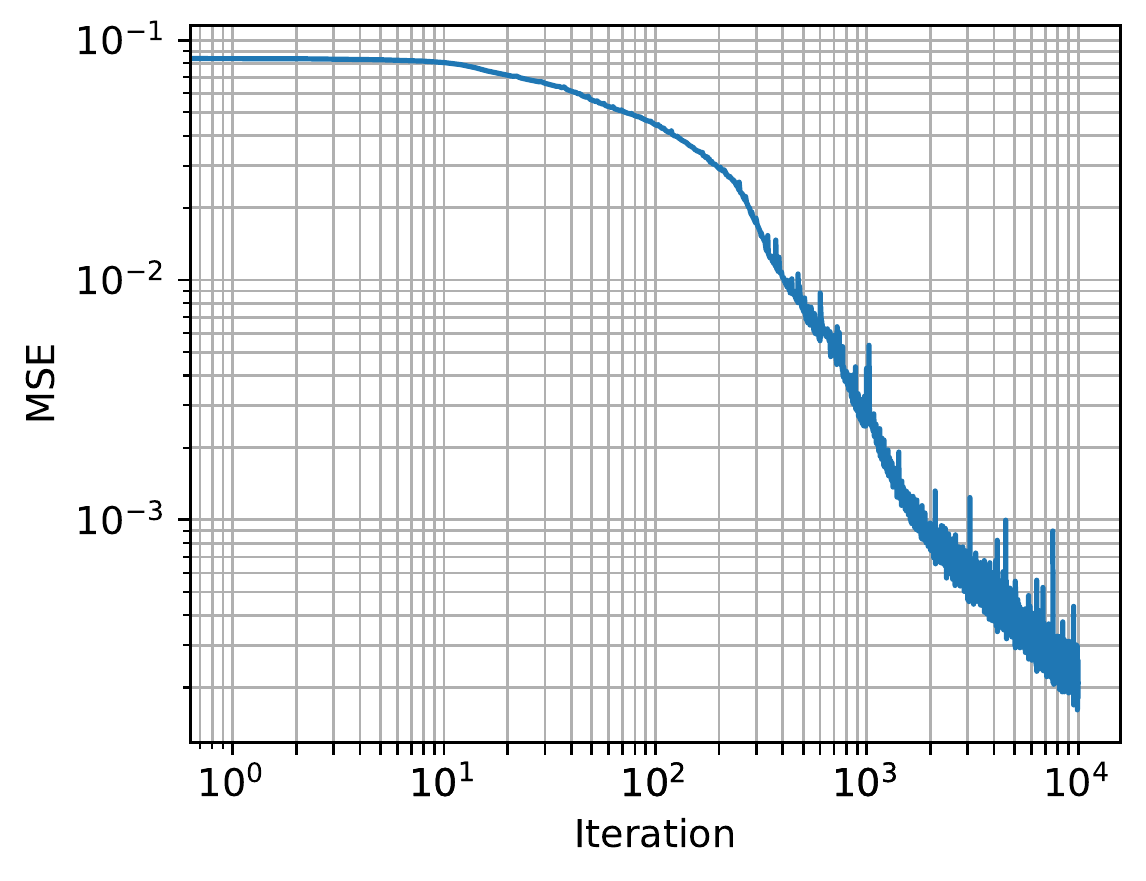}\label{fig:model_training}}
\subfloat[Quantum]{\includegraphics[scale=0.75]{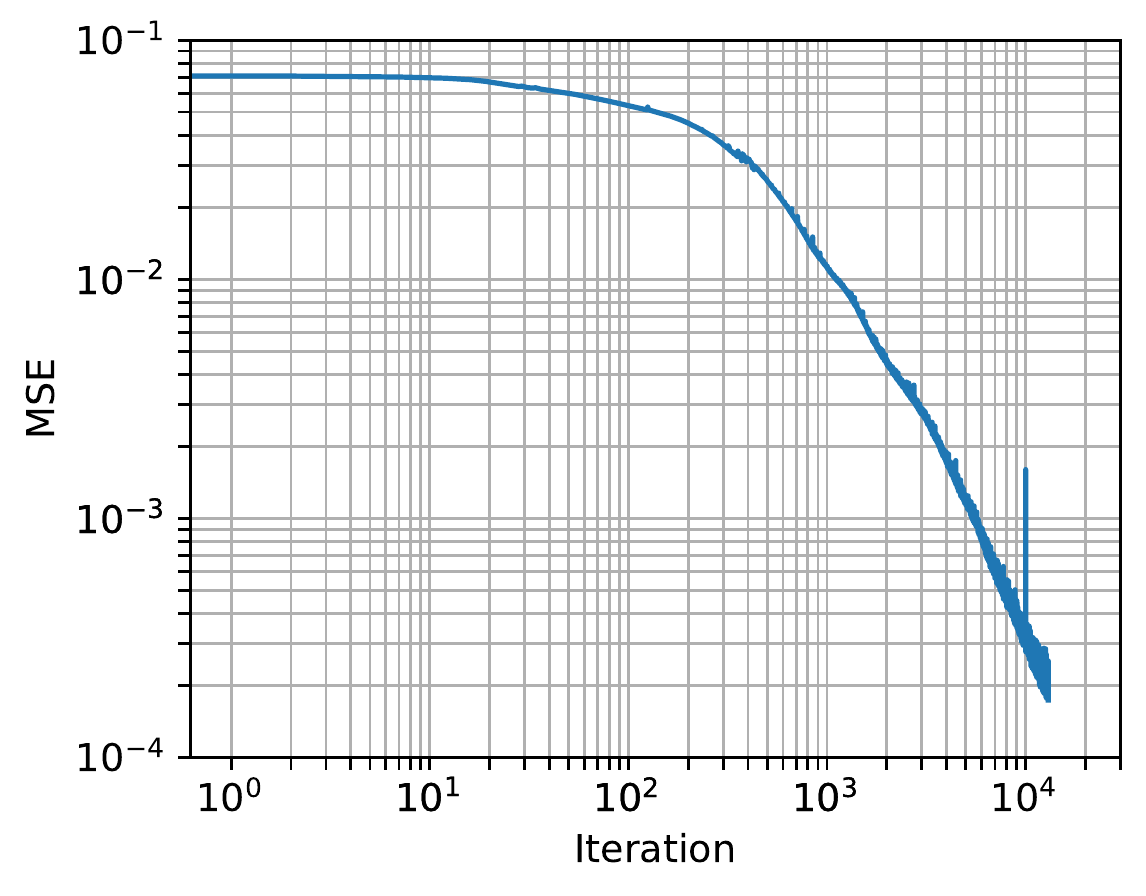}\label{fig:model_complex_training}}
\caption{The proposed machine learning structure is trained over a set consisting of 3500 examples. The set consists of random control pulses and the corresponding simulated output optical power for the classical application, and interferometer power measurements for the fully-quantum application. The plot shows the MSE, which is used as the loss function for the training, versus the number of iterations for the (a) classical application and (b) the quantum application.}
\end{figure*}

For the task of modeling the chip, the MSE obtained after $10^4$ iterations was about $2.1\times10^{-4}$ for the training dataset. Figure \ref{fig:model_training} shows the MSE versus the number of iterations. For the testing dataset, the MSE evaluated is $3.4 \times10^{-4}$. Supplementary Figures \ref{fig:ex_0},\ref{fig:ex_1}, and \ref{fig:ex_2} show examples selected randomly of the testing dataset including the control voltages, simulated measured waveguide power distribution and the predicted power distribution.

To test the control part, we defined as an example a sequence of target unitaries in the time interval $0 \le t \le 300 (ms)$, given by 
\begin{align}
	U(t) = \begin{cases}
		X_{13} & 50 \le t < 80 \\
		H_{13} & \left(110 \le t < 140\right) \lor \left(250 \le t < 280\right)\\
		X_{12} & 170 \le t < 210 \\
		I          & \text{otherwise} 
	\end{cases}
\label{equ:seq}	
\end{align}
where the unitaries are defined in Table \ref{tab:targets}. The Hamiltonian is then evaluated for each time interval by taking the matrix logarithm $H = \frac{i}{l}\log U$.

After training the controller model for $500$ iterations, the MSE was $2 \times 10^{-2}$. The MSE versus the number of iterations is plotted in Figure \ref{fig:controller_training}. The resulting control voltages are shown in Supplementary Figure \ref{fig:controller_voltages}, and the resulting predicted ideal power distribution in Supplementary Figure \ref{fig:controller_powers}.
\begin{table}[t]
\centering
\caption{Target Hamiltonians used for testing the proposed architecture for the controller as per Equation \ref{equ:seq} for the classical model, and Equation \ref{equ:seq_complex} for the fully-quantum model.}
\label{tab:targets}
\begin{tabular}{|c|c|c|}
\hline
\textbf{Symbol}   & \textbf{Expression} & \textbf{Description} \\ \hline
$I$       & $\begin{pmatrix} 1 & 0 & 0 \\ 0 & 1 & 0 \\ 0 & 0 & 1\end{pmatrix}$  & Identity (100\% decoupling between waveguides) \\ \hline
$X_{13}$       & $\begin{pmatrix} 0 & 0 & 1 \\ 0 & 1 & 0 \\ 1 & 0 & 0\end{pmatrix}$  & Perfect Transfer between waveguide 1 and waveguide 3 \\ \hline
$H_{13}$       & $\begin{pmatrix} \frac{1}{\sqrt{2}} & 0 & \frac{1}{\sqrt{2}} \\ 0 & 1 & 0 \\ \frac{1}{\sqrt{2}} & 0 & \frac{-1}{\sqrt{2}}\end{pmatrix}$  & 50-50 Power split between waveguide 1 and waveguide 3 (Hadamard gate) \\ \hline
$X_{12} $       & $\begin{pmatrix} 0 & 1 & 0 \\ 1 & 0 & 0 \\ 0 & 0 & 1\end{pmatrix}$  & Perfect transfer between waveguide 1 and waveguide 2 \\ \hline
$Z_{13} $       & $\begin{pmatrix} 1 & 0 & 0 \\ 0 & 1 & 0 \\ 0 & 0 & -1\end{pmatrix}$  & Phase shift of $\pi$ between waveguide 1 and waveguide 3 \\ \hline
$R_{X_{13}}(\theta) $       & $\exp{\left(-i\theta X_{13}\right)}$  & Rotation about X-axis by angle $\theta$ between waveguide 1 and waveguide 3\\ \hline
 $R_{Z_{13}}(\theta) $       & $\exp{\left(-i\theta Z_{13}\right)}$  & Rotation about Z-axis by angle $\theta$ between waveguide 1 and waveguide 3\\ \hline
\end{tabular}
\end{table}

For the second application which is the fully-quantum setting, we use the same dataset of pulses, but now we have the interferometer power measurements as the model output. The number of iterations is $1.3\times 10^4$, which is more than the other model to account for doubling the size of the outputs. Figure \ref{fig:model_complex_training} shows the performance of the training in this case. The MSE evaluated for the testing dataset it $2.88\times10^{-4}$, while it was $1.74 \times 10^{-4}$ for the training set. This is an indication for the the ability of the model to fit the training dataset as well as generalize to the testing dataset. Supplementary Figures \ref{fig:ex_complex_0} and \ref{fig:ex_complex_1} show the result of the predicted waveforms using the same control pulses as in Supplementary Figures \ref{fig:ex_0} and \ref{fig:ex_1}. Now, since the phase is also measured, then we can have a complete quantum description of the output state, and thus we can construct the evolution unitary. A commonly used measure for the closeness of two quantum gates $U$ and $V$ of dimension $d$, is the gate infidelity defined as 
\begin{equation}
	1-F(U,V) = 1-\frac{|\tr{\left(U^{\dagger}V\right)}|^2}{d^2}.
\end{equation}
Infidelity is thus a number between 0 and 1, with 0 representing complete overlap (i.e. same matrices). Supplementary Figure \ref{fig:infidelity_model} shows the infidelity between the predicted unitary and actual unitary as a function of time for these two examples. Finally, for evaluating the control algorithm in this setting, we used as an example the following sequence for $0<t<280 (ms)$
\begin{align}
	U(t) = \begin{cases}
		X_{13} & 50 \le t < 90 \\
		R_{X_{13}}\left(\frac{\pi}{4}\right) & 130 \le t < 170  \\
		R_{Z_{13}}\left(0.1\right) & 210 \le t < 250 \\
		I          & \text{otherwise} 
	\end{cases}
\label{equ:seq_complex}	
\end{align}
The history of the MSE of the controller during the training is shown in Figure \ref{fig:controller_complex_training}. The resulting infidelity between the desired sequence of quantum gates and the controlled quantum gates are shown in Figure \ref{fig:controller_complex_infidelity}, while the control voltages are shown in Supplementary Figure \ref{fig:controller_complex_voltages}. 

\begin{figure*}[h]
\centering
\subfloat[Classical]{\includegraphics[scale=0.75]{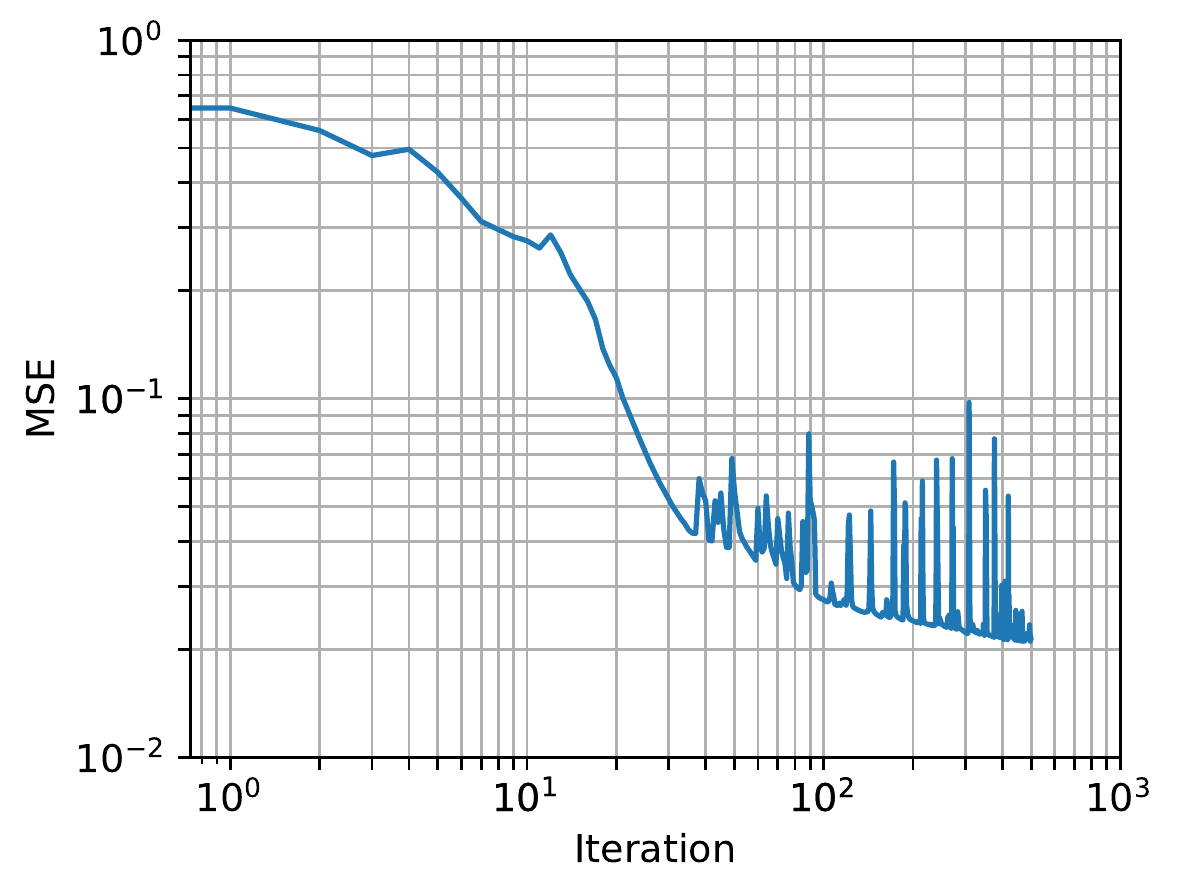}\label{fig:controller_training}}
\subfloat[Quantum]{\includegraphics[scale=0.75]{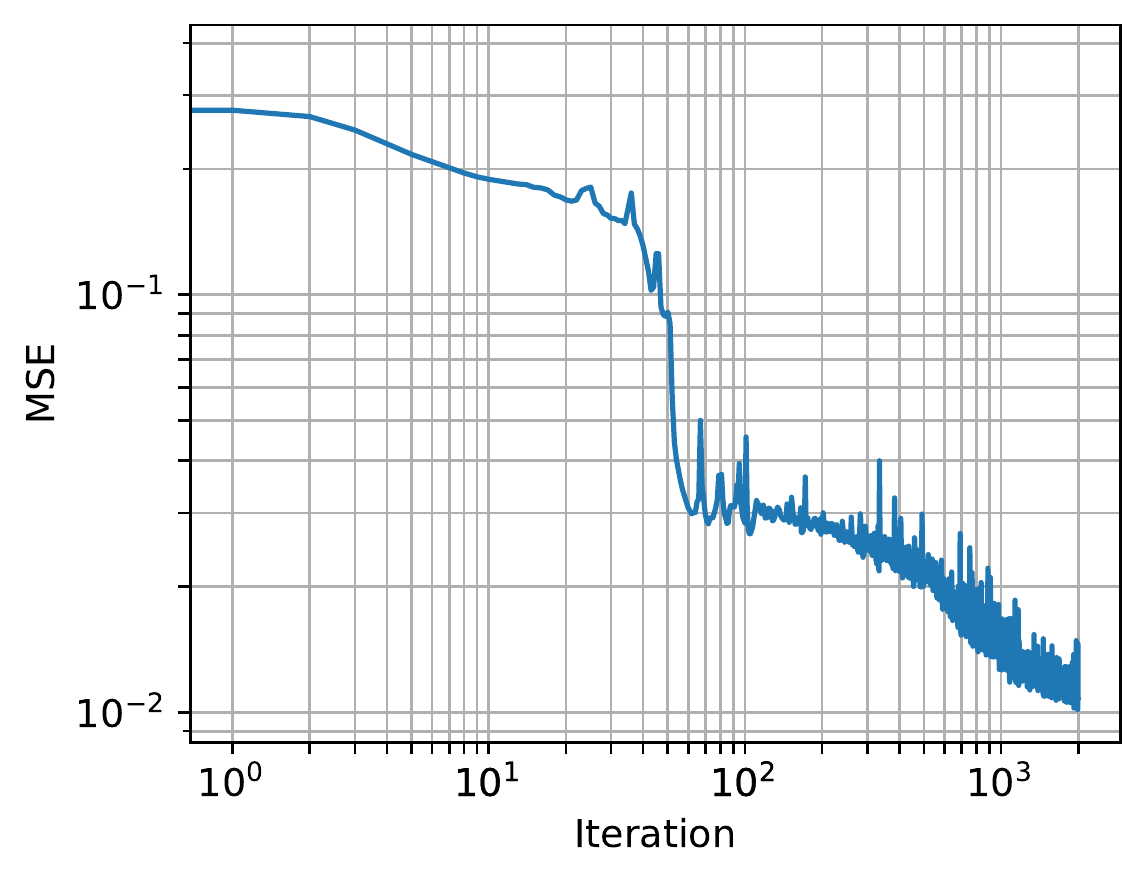}\label{fig:controller_complex_training}}
\caption {The proposed controller architecture is trained to obtain the control voltages needed to achieve the target sequence of (a) optical output power distribution for the classical application defined in Equation \ref{equ:seq}, and (b) unitary gates for the quantum model defined in Equation \ref{equ:seq_complex}. The plot shows the MSE used as the loss function versus the number of iterations.}
\end{figure*}

\begin{figure*}[!ht]
\centering
\includegraphics[scale=0.75]{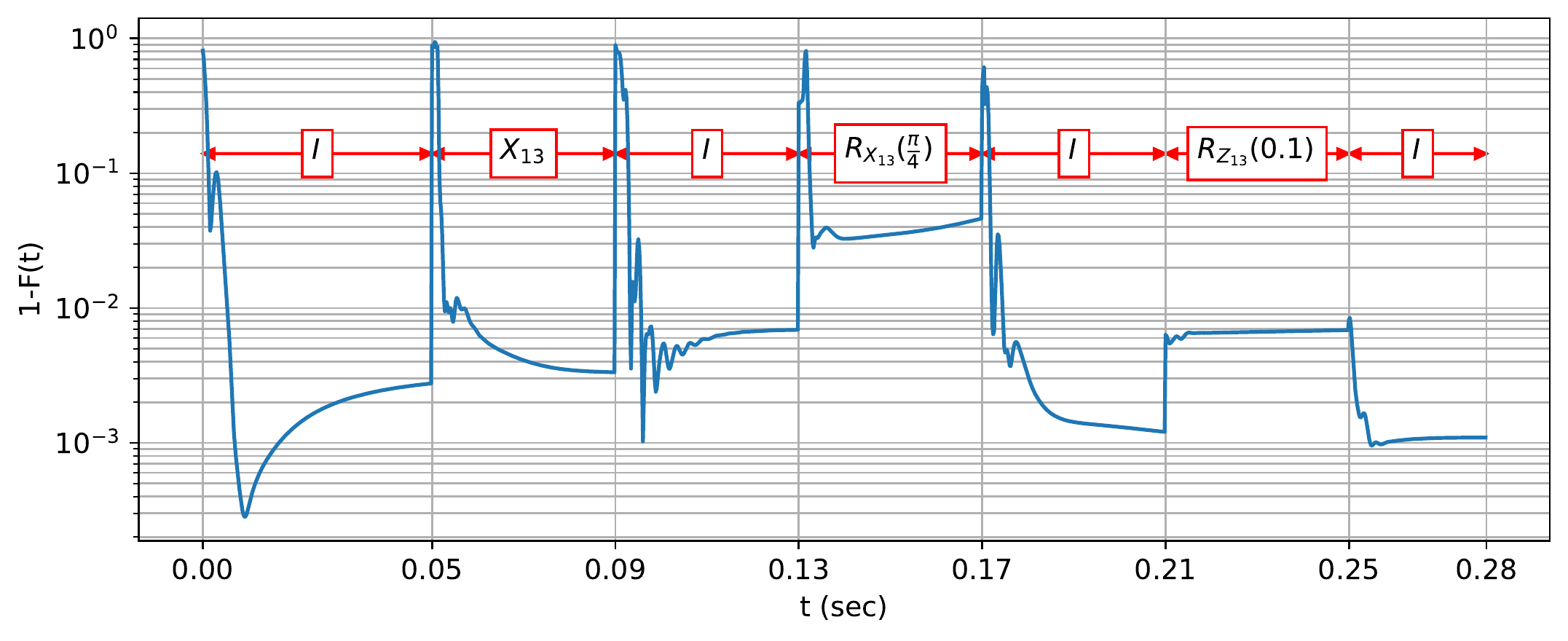}
\caption {The resulting infidelity for the sequence of target quantum gates defined in Equation \ref{equ:seq_complex}, after training the proposed controller architecture. Photons arriving at different time instants would see different quantum gates. The plot shows the infidelity of between the ideal quantum gate and the actual gate obtained by the controller at each time step.}
\label{fig:controller_complex_infidelity}
\end{figure*}

\subsection{Discussion}
The presented results show the accuracy of the proposed architecture in modeling the chip with all the constraints mentioned earlier. Quantitatively, the loss represented by the MSE decreases on average by increasing the number of iterations during the training phase, reaching a small value that is in order of $10^{-4}$. However, this is not sufficient to completely asses the behavior of the proposed algorithm. The plots of the waveforms in Supplementary Figures \ref{fig:ex_0},\ref{fig:ex_1}, and \ref{fig:ex_2} show qualitatively the accuracy of the model. The difference between the predicted and simulated power distribution is almost negligible. More importantly, since the model has not been trained on the testing set, it proves that the proposed structure can generalize. This important for the task of modeling. The architecture does not allow to give explicit mathematical expression for the Hamiltonian. But, due to its ability to generalize, we can just use it directly to estimate the Hamiltonian given the control voltages. Also, quantitatively the MSE evaluated for the testing set is also in the order of $10^{-4}$, without much degradation than the value for the training set. 

The qualitative results also show that the architecture is able to handle all the challenges described in Section \ref{sec:challenges}. We were able to model the distortions caused by the equivalent circuit without the need to explicitly define a particular circuit model or how the Hamiltonian depends on the circuit response. This also saves us from having to characterize these parasitic effects experimentally, which is difficult as discussed previously.

For the control task, the proposed method was also very successful in obtaining the required control voltages as reflected in Supplementary Figure \ref{fig:controller_powers}. We see that the distortions that were present in the power distributions are not there anymore, and at the same time we were able to achieve the required functionality. The control voltages were also limited to the desired operating range. However, we see that for the $X_{12}$ gate, the algorithm could not do full transfer between waveguides 1 and 2. We believe that this is related to the fact that not all gates are possible to implement, which is a subject of the future work. A final thing to notice is that all the examples in the training set were limited to the time range $0 \le t \le 200 (ms)$. However, the target control sequence has a wider range  $0 \le t \le 300 (ms)$, and still we are successful in our task. This is a result of using the GRU layers, and shows how the whole model generalizes quite well. 

The proposed modifications in the architecture to account for fully-quantum models was also very successful. This is evident from the low MSE value for both training and testing datasets with small difference between both. This is supported qualitatively through the plots of the power waveforms and infidelity versus time. Also, the controller architecture seems to perform quite well. The example we tested shows the possibility of implementing some basic quantum gates which are identity, Pauli X,  rotation about X-axis with angle $\pi/4$ which is equivalent to a Hadamard gate with phase shifts, and rotation about Z-axis. At each time instant, the photon traveling through the chip will sense a different quantum gate. The gate infidelities at all time instants, apart from the transition moments, are low (worst case was $4\times10^{-2}$). The gates act on a qubit spatially encoded between the first and last waveguide. However, there is a major advantage for our proposed controller architecture, which is the input is the target sequence of quantum gates rather than a single gate. In general, the control voltages required for realizing a particular gate can depend on the previous history of gates realized so far due to the drifting problem described earlier. In other words, the same gate could need different control pulses at different points in time during the operation of the chip. Our proposed method deals automatically with this issue compared to standard quantum control literature that deals with one target quantum gate only \cite{khaneja2005optimal, caneva2011chopped, machnes2015gradient}. 
\section{Conclusion}\label{sec:conclusions}
In this paper, we proposed a deep learning structure that is suitable to model a reconfigurable integrated waveguide array chip. The architecture addresses three major problems faced when characterizing the chip experimentally. The uncertainty in the Hamiltonian model, the presence of undesired macroscopic dynamics causing distortions, and losses due to imperfect measurements. The proposed architecture followed a graybox model approach, where the Hamiltonian as a function of control voltages is treated as a blackbox utilizing a GRU network as a main component. The waveguide power distribution as function of the Hamiltonian is treated as a whitebox since the laws of quantum mechanics are known. We also proposed another complementary deep learning structure to obtain the control voltages required to achieve some target sequence of gates. The qualitative as well as quantitative results showed a very promising performance for both tasks.   

There are many possible extensions to the presented work. On the theoretical side, it would be interesting to know the set of gates that are possible to implement on this chip given the constraints introduced in the model, such as limited control voltages. We would like also to validate the numerical results shown in the paper experimentally on the physical chip which is currently in progress. Another interesting extension is to explore the use of fidelity as cost function to do the training rather than the MSE, and see whether or not would it yield better results. Finally, it would be worth looking into extending the methods introduced in this paper to model and control other quantum systems. 

\paragraph*{Acknowledgments:}
AY is supported by an Australian Government Research Training Program Scholarship, and acknowledges RMIT University for hosting him during his visit. A.P. acknowledges funding from the Australian Research Council Centre for Quantum Computation and Communication Technology CE170100012; Australian Research Council Discovery Early Career Researcher Award, Project No. DE140101700; RMIT University Vice-Chancellors Senior Research Fellowship and a Google Faculty Research Award.
MT and CF acknowledge Australian Research Council Discovery Early Career Researcher Awards, projects No.\ DE160100821 and DE170100421, respectively. 
This research is also supported in part by the ARCLab facility at UTS.
\bibliographystyle{alphaurl}
\bibliography{library} 
\newpage
\appendix
\section{Supplementary figures}\label{sec:suppfig}

\begin{figure*}[h]
\centering
\includegraphics*[scale=0.75]{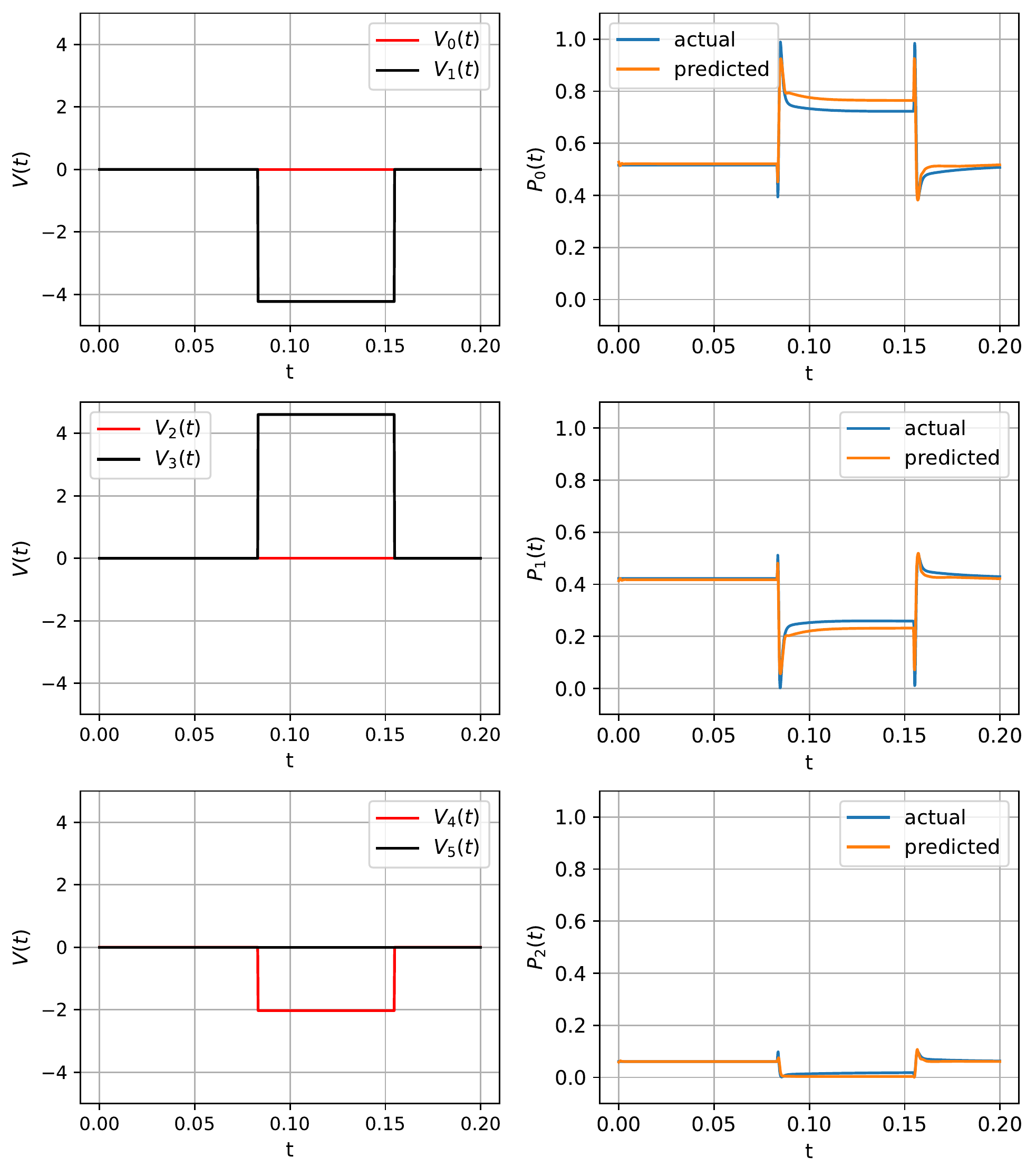}
\caption {Random example from the testing dataset. The left column is the control voltages applied to the electrodes across the waveguides. The right column is the waveguide output power distribution, both simulated as in the dataset and predicted by the proposed algorithm. The initial state is $\ket{0}$, i.e. full power at the first waveguide.}
\label{fig:ex_0}
\end{figure*}

\begin{figure*}[h]
\centering
\includegraphics[scale=0.75]{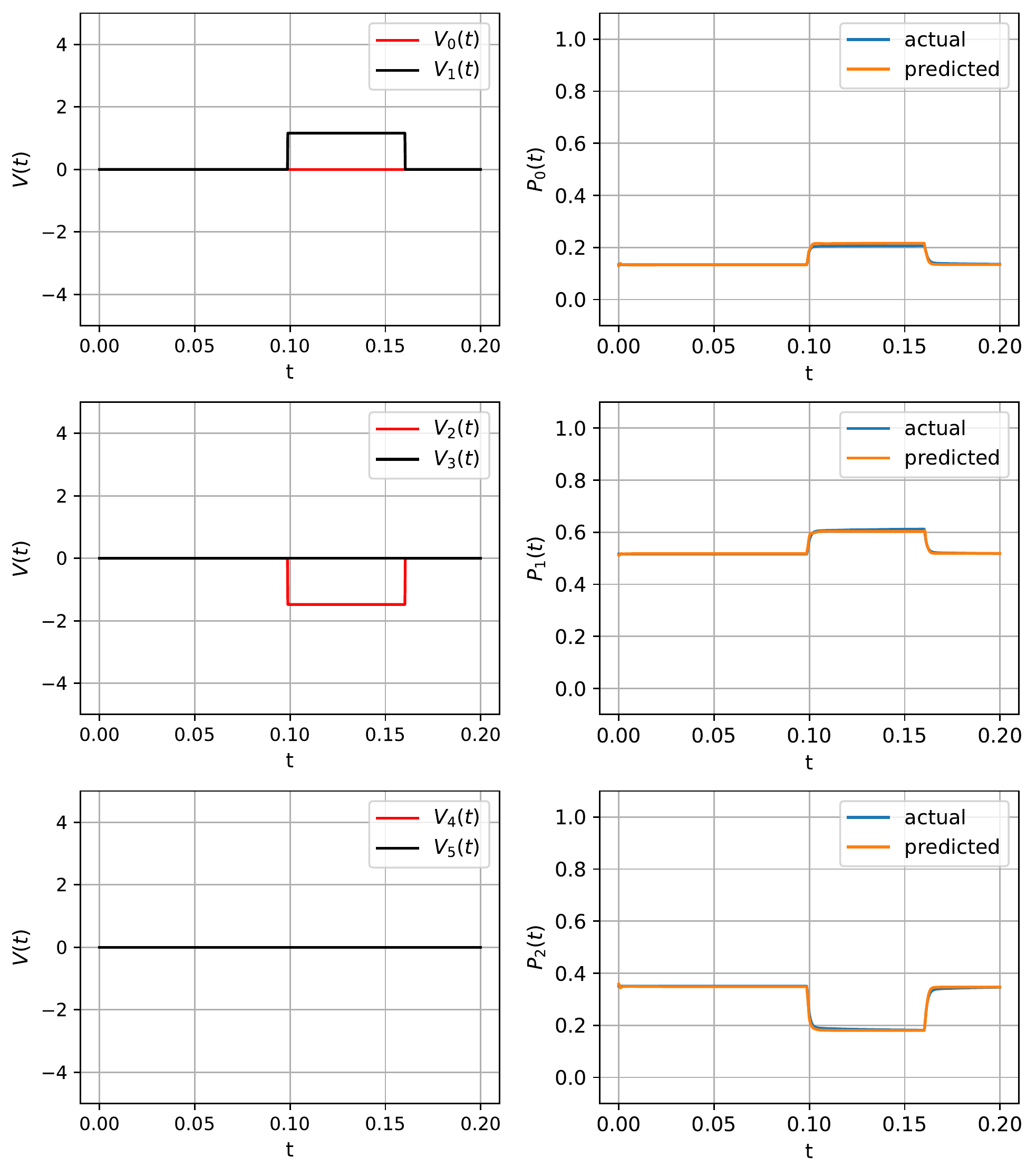}
\caption {Random example from the testing dataset. The left column is the control voltages applied to the electrodes across the waveguides. The right column is the waveguide output power distribution, both simulated as in the dataset and predicted by the proposed algorithm. The initial state is $\ket{2}$, i.e. full power at the third waveguide.}
\label{fig:ex_1}
\end{figure*}

\begin{figure*}[h]
\centering
\includegraphics[scale=0.75]{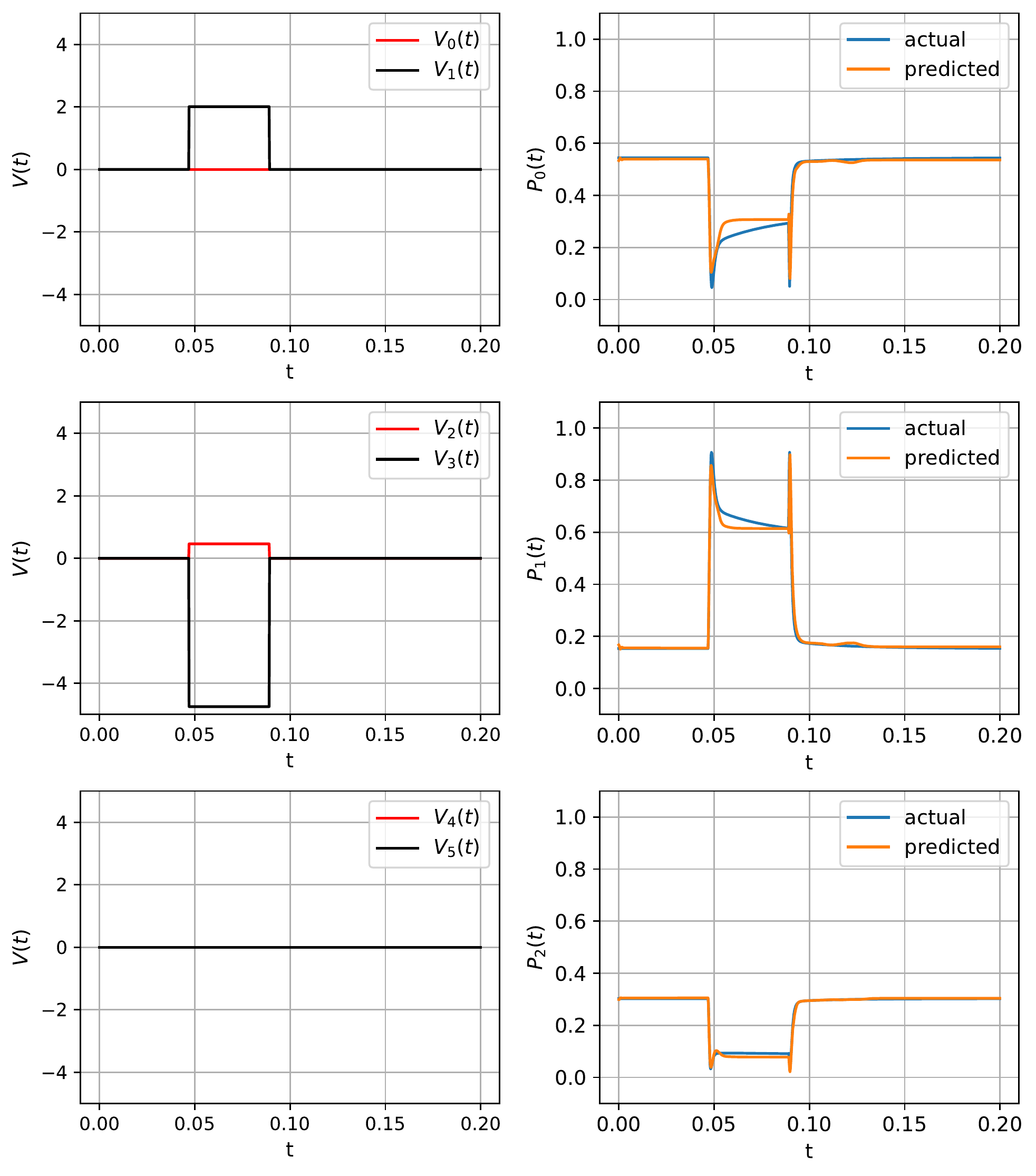}
\caption {Random example from the testing dataset. The left column is the control voltages applied to the electrodes across the waveguides. The right column is the waveguide output power distribution, both simulated as in the dataset and predicted by the proposed algorithm. The initial state is $\ket{1}$, i.e. full power at the second waveguide.}
\label{fig:ex_2}
\end{figure*}

\begin{figure*}[h]
\centering
\includegraphics[scale=0.75]{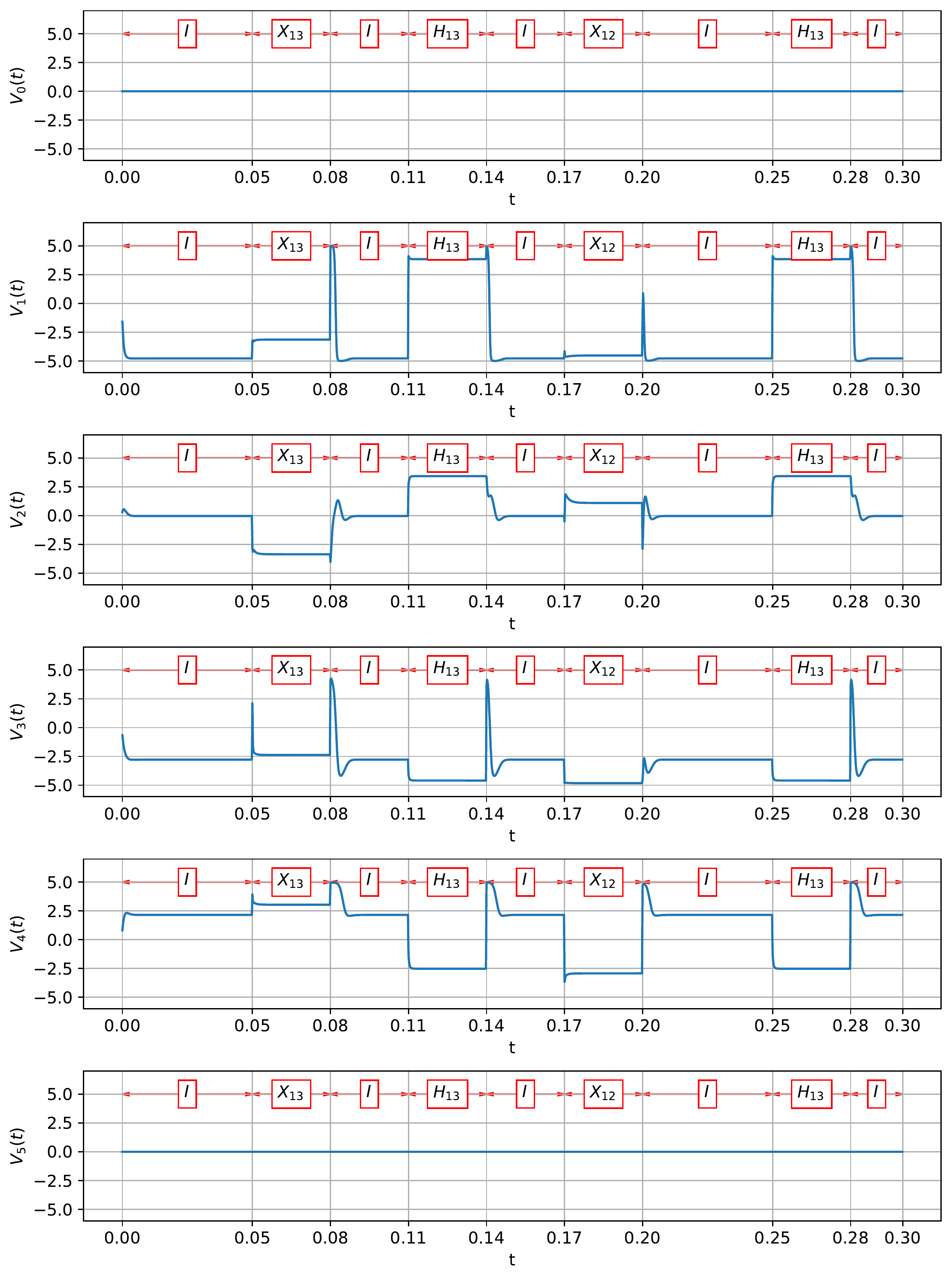}
\caption {The resulting control voltages to obtain the sequence of targets defined in Equation~\ref{equ:seq}.}
\label{fig:controller_voltages}
\end{figure*}

\begin{figure*}[h]
\centering
\subfloat[initial state = $\ket{0}$]{\includegraphics[scale=0.5]{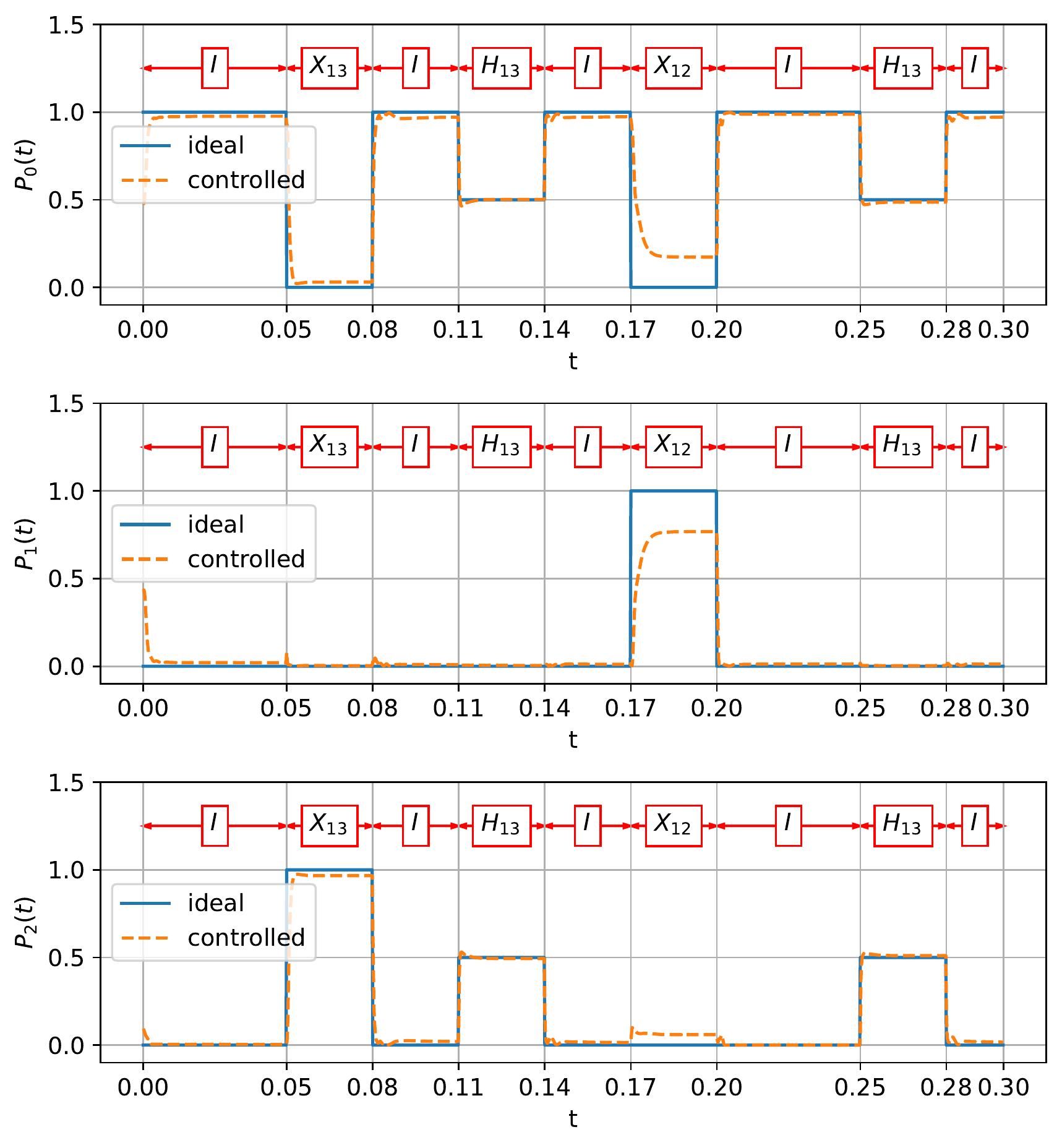}}
\subfloat[initial state = $\ket{1}$]{\includegraphics[scale=0.5]{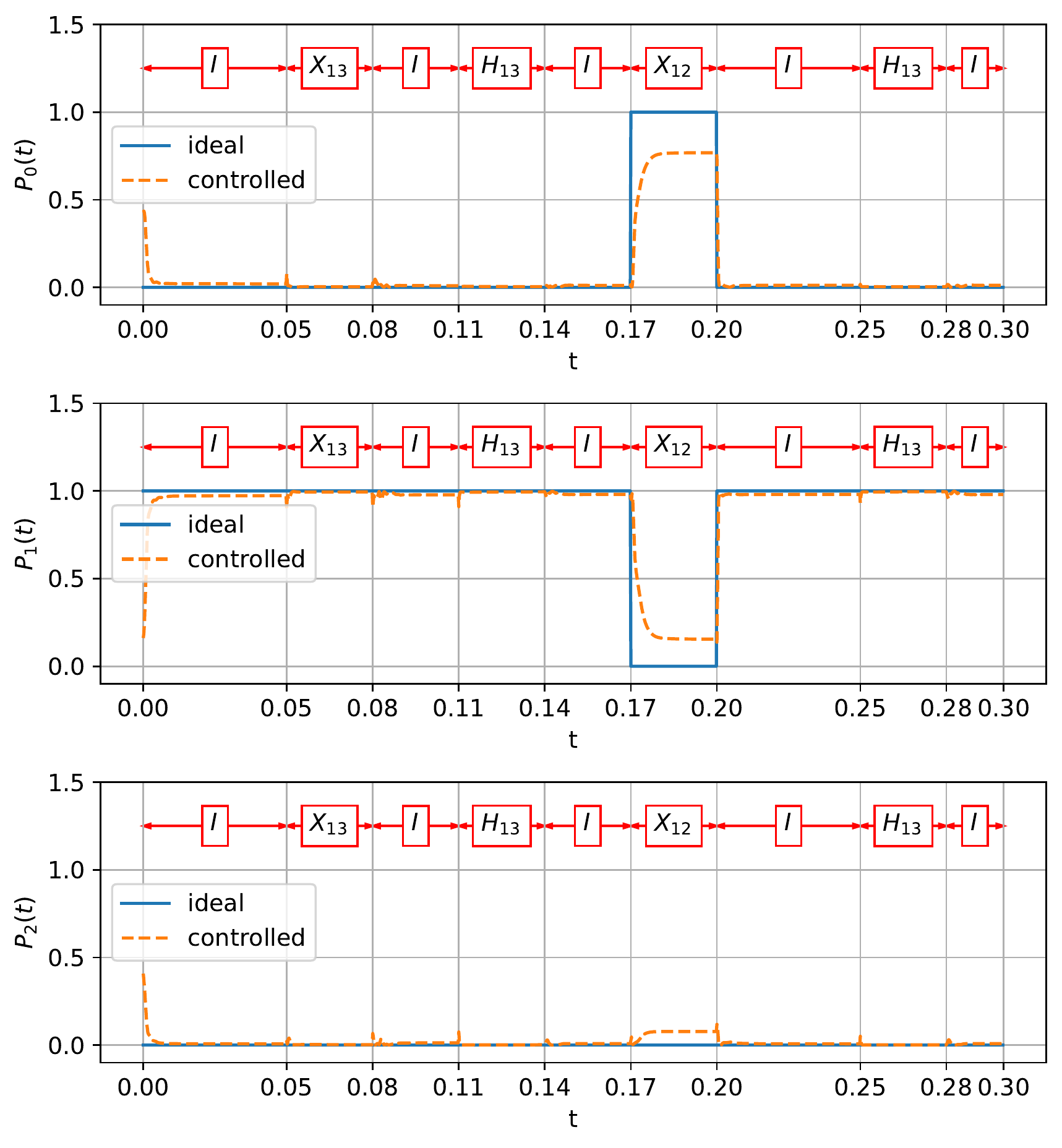}}\\
\subfloat[initial state = $\ket{2}$]{\includegraphics[scale=0.5]{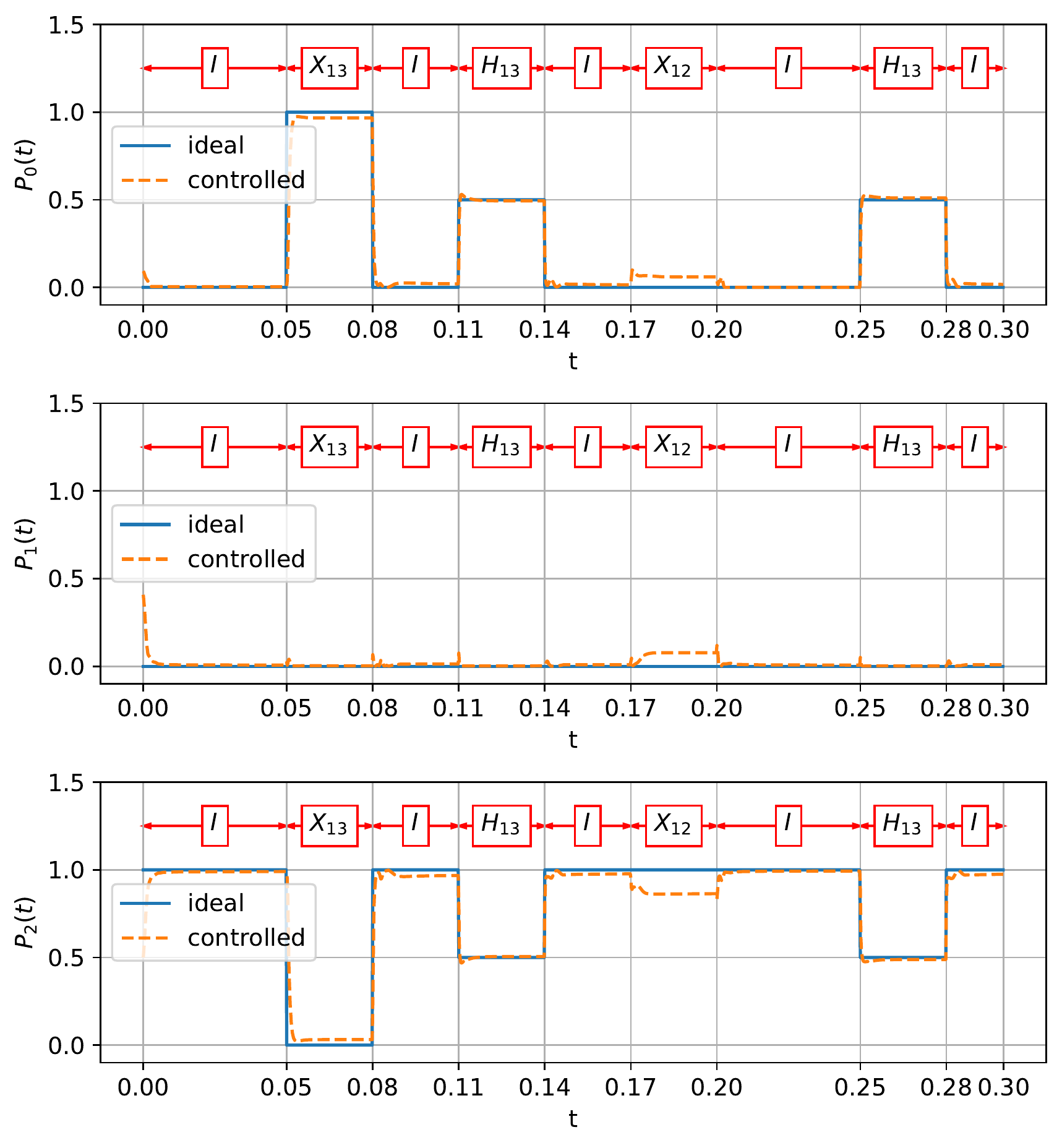}}
\caption {The resulting waveguide distribution realizing the sequence of targets defined in Equation \ref{equ:seq}, for the initial states a) $\ket{0}$, b) $\ket{1}$, and c) $\ket{2}$.}
\label{fig:controller_powers}
\end{figure*}

\begin{figure}[h]
\centering
\includegraphics[scale=0.75]{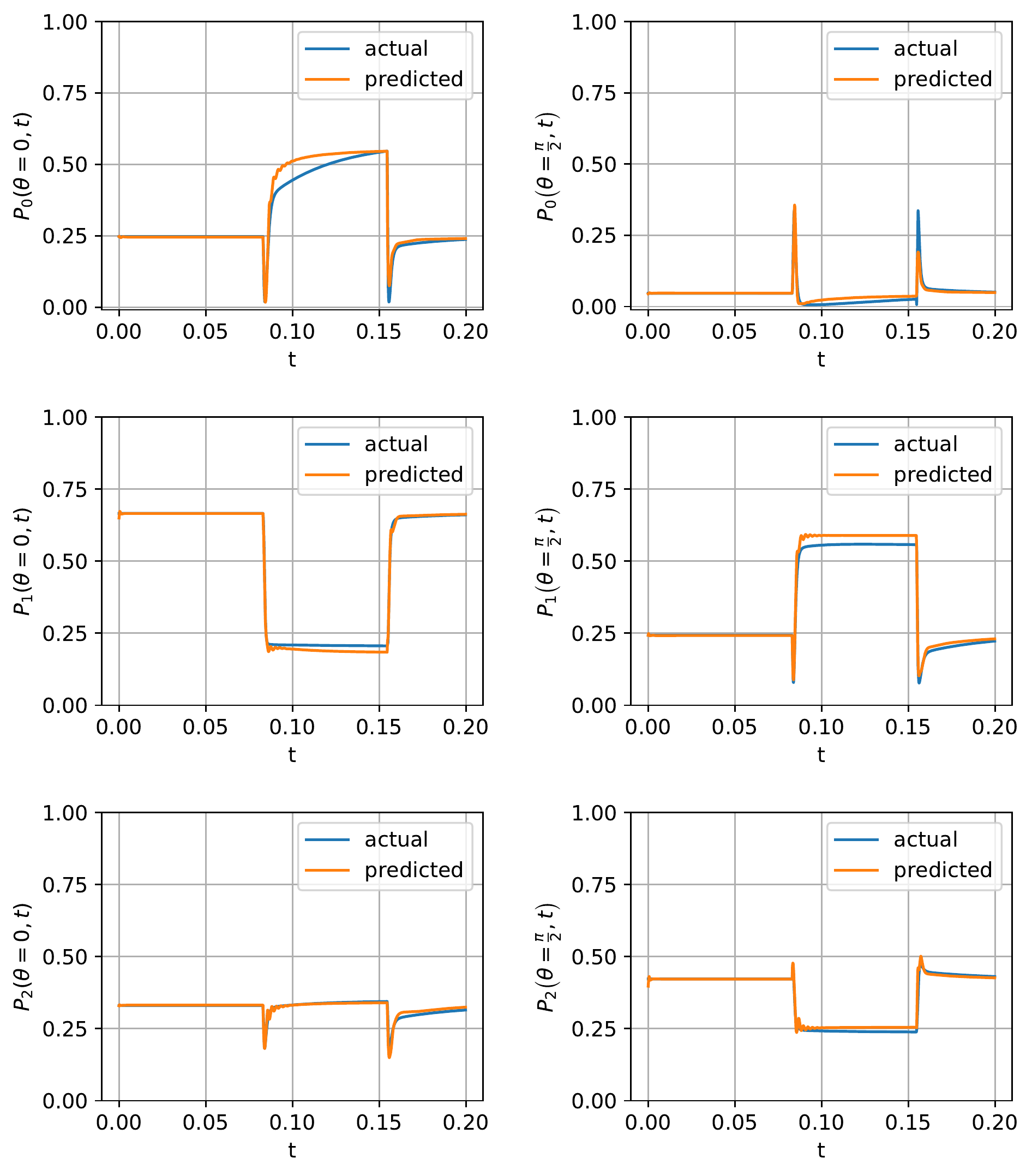}
\caption {Random example from the testing dataset showing the interferometer power measurements across each waveguide, both simulated as in the dataset and predicted by the proposed algorithm. The same control voltages are applied as in Supplementary Figure \ref{fig:ex_0}. The initial state is $\ket{0}$.}
\label{fig:ex_complex_0}
\end{figure}

\begin{figure}[h]
\centering
\includegraphics[scale=0.75]{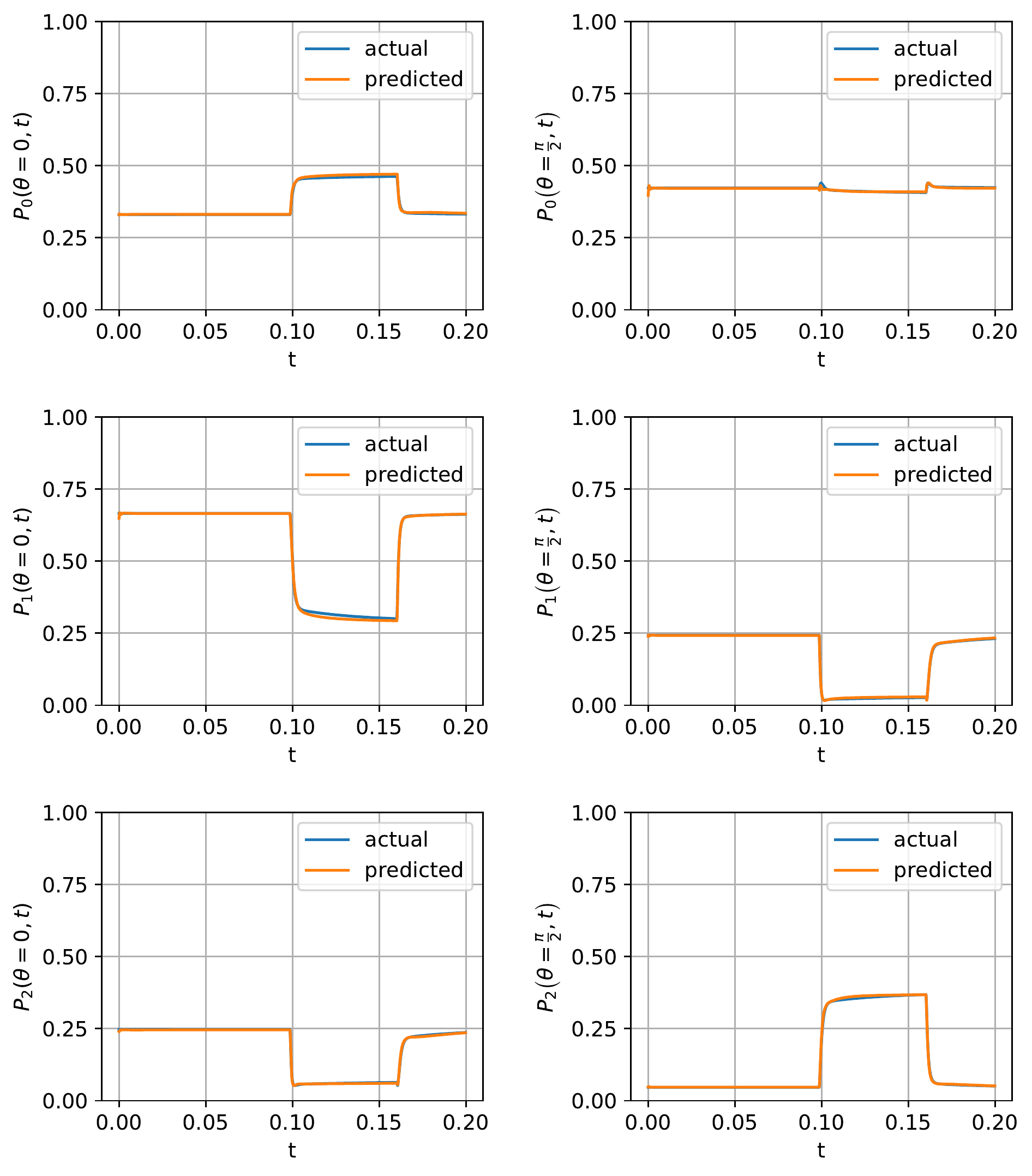}
\caption {Random example from the testing dataset showing the interferometer power measurements across each waveguide, both simulated as in the dataset and predicted by the proposed algorithm. The same control voltages are applied as in Supplementary Figure \ref{fig:ex_1}. The initial state is $\ket{2}$.}
\label{fig:ex_complex_1}
\end{figure}

\begin{figure}[h]
\centering
\subfloat[Infidelity for the example in Supplementary Figure \ref{fig:ex_complex_0}]{\includegraphics[scale=0.75]{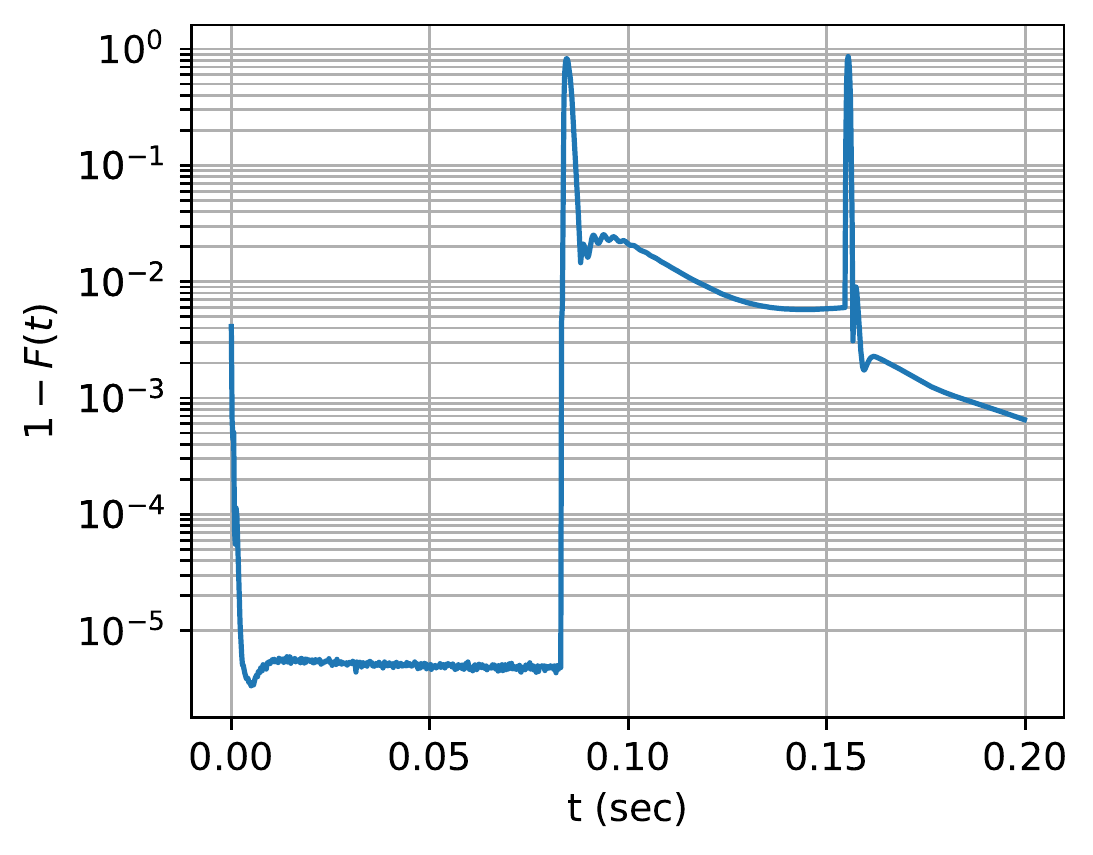}}
\subfloat[Infidelity for the example in Supplementary  Figure \ref{fig:ex_complex_1}]{\includegraphics[scale=0.75]{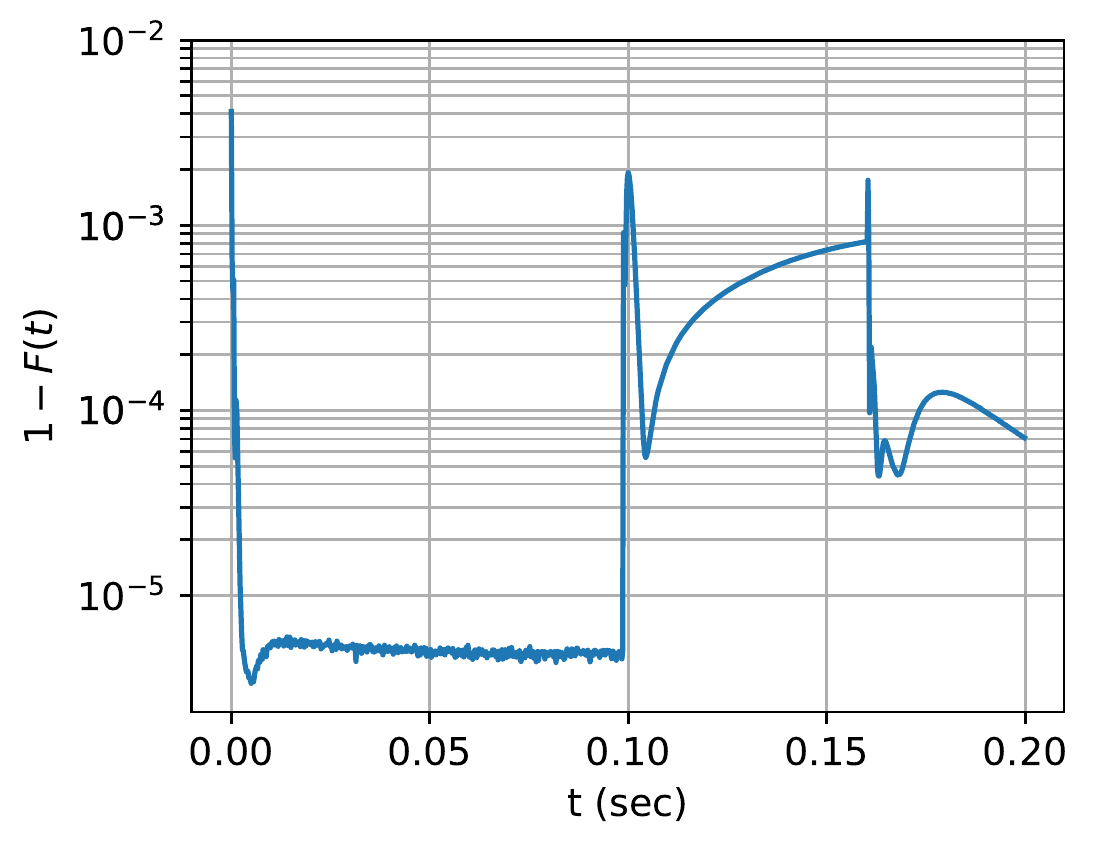}}
\caption{The gate infidelity evaluated between the predicted evolution unitary and the actual evolution unitary for two random examples from the testing set. }
\label{fig:infidelity_model}
\end{figure}

\begin{figure*}[h]
\centering
\includegraphics[scale=0.75]{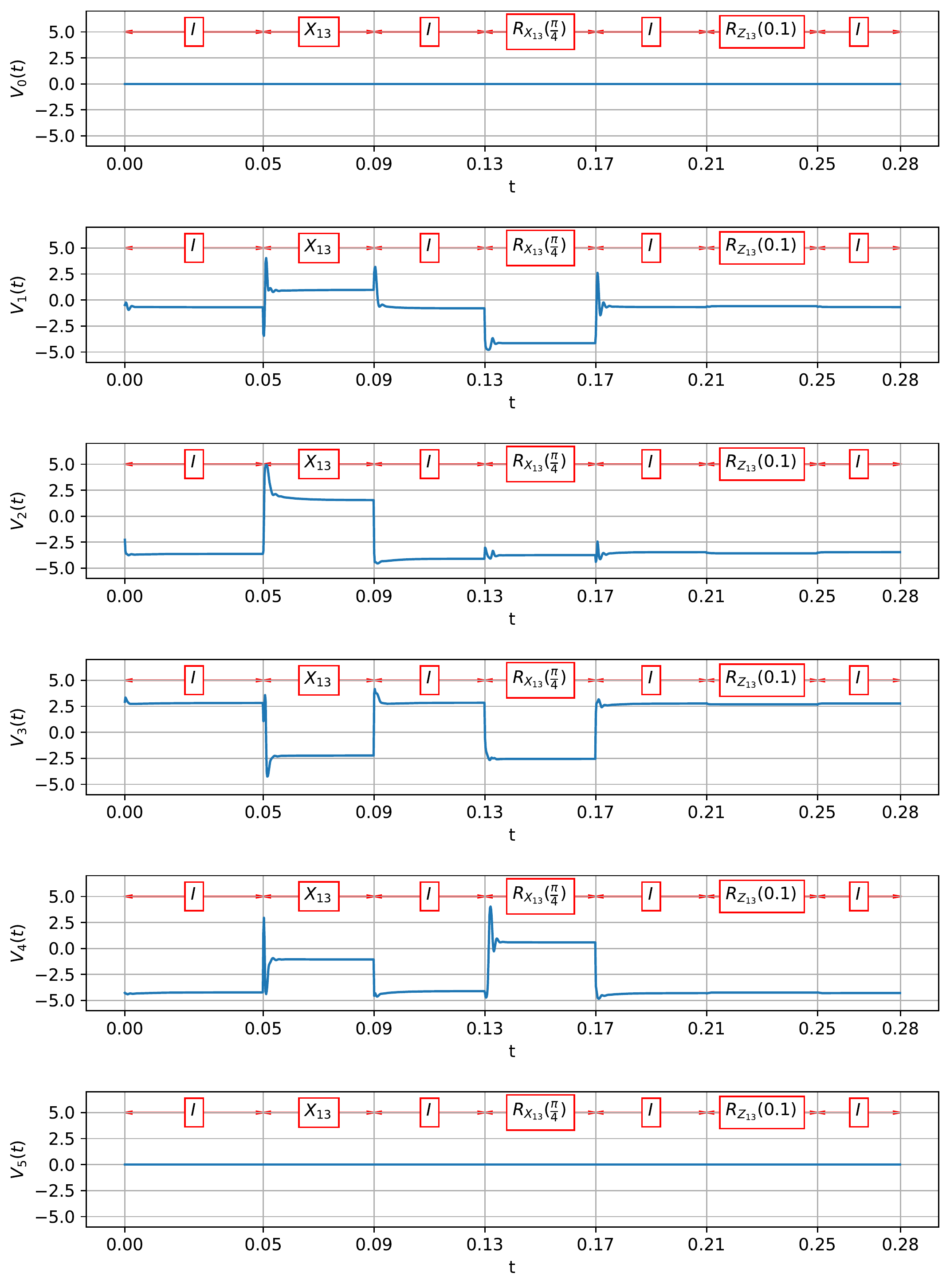}
\caption {The resulting control voltages to obtain the sequence of targets defined in Equation \ref{equ:seq_complex}.}
\label{fig:controller_complex_voltages}
\end{figure*}
\end{document}